\documentclass[a4paper,fleqn,usenatbib]{mnras}

\usepackage[T1]{fontenc}
\usepackage{ae,aecompl}

\usepackage{graphicx}	
\usepackage{amsmath}	
\usepackage{amssymb}	
\usepackage{aas_macros}
\usepackage{color}

\title[Mass segregation with galaxy analogues]{Uncovering Mass Segregation with Galaxy Analogues in Dark Matter Simulations}

\author[G. D. Joshi, L. C. Parker and J. Wadsley]{Gandhali D. Joshi,\thanks{Email: joshigd@mcmaster.ca} Laura C. Parker, James Wadsley \\
Department of Physics and Astronomy, McMaster University, Hamilton, ON L8S 4M1, Canada}

\date{Accepted XXX. Received YYY; in original form ZZZ}

\pubyear{2016}

\begin{document}
\label{firstpage}
\pagerange{\pageref{firstpage}--\pageref{lastpage}}
\maketitle

\begin{abstract}
We investigate mass segregation in group and cluster environments by identifying galaxy analogues in high-resolution dark matter simulations. Subhalos identified by the AHF and ROCKSTAR halo finders have similar mass functions, independent of resolution, but different radial distributions due to significantly different subhalo hierarchies. We propose a simple way to classify subhalos as galaxy analogues. The radial distributions of galaxy analogues agree well at large halo-centric radii for both AHF and ROCKSTAR but disagree near parent halo centres where the phase-space information used by ROCKSTAR is essential.

We see clear mass segregation at small radii (within $0.5\,r_{vir}$) with average galaxy analogue mass decreasing with radius. Beyond the virial radius, we find a mild trend where the average galaxy analogue mass increases with radius. These mass segregation trends are strongest in small groups and dominated by the segregation of low mass analogues. The lack of mass segregation in massive galaxy analogues suggests that the observed trends are driven by the complex accretion histories of the parent halos rather than dynamical friction.
\end{abstract}




\section{Introduction}
Galaxies in groups and clusters are known to exhibit different properties compared to field galaxies. They have redder colours, more elliptical morphologies and suppressed star formation rates \citep[e.g.][]{Oemler74,Dressler80,Balogh04,Hogg04,Kauffmann04,Blanton05}. On large scales, structure in the Universe grows hierarchically; smaller dark matter halos collapse earlier while larger structures form later through the coalescence of these smaller halos. Baryons are accreted into the potentials of these halos to form galaxies. As they are accreted onto larger objects, their properties transition from those of field galaxies to those of group/cluster galaxies. This is evidenced by observed radial trends in several different properties such as luminosity and morphology \citep[e.g.][]{Girardi03}, colour \citep[e.g.][]{Blanton07}, quenched fractions \citep[e.g.][]{Wetzel12} and star formation rates \citep[e.g.][]{Balogh00}. Such a correlation between an average galaxy property and distance from the group/cluster centre is defined as segregation.

Segregation in observed properties may be largely driven by baryonic physics. Several mechanisms can transform galaxies in groups/clusters - harassment and tidal interactions with other nearby galaxies can remove gas, stars and dark matter \citep{Moore96,Moore98}; gas removal can result in strangulation and preventing future star formation \citep{Larson80,Balogh00,Kawata08}; ram pressure stripping can remove the more bound cold gas \citep{Gunn72,Abadi99}; mergers can trigger starbursts that rapidly consume the fuel for star formation \citep{Makino97,Angulo09}. Mass segregation, on the other hand, could arise due purely to the interactions of the dark matter halos in which galaxies reside and the larger potential of the group or cluster. Understanding mass segregation may shed light on the processes of galaxy evolution in these environments and whether baryons or dark matter play the dominant role. Additionally, several galaxy properties such as luminosity, stellar ages and metallicities etc. are correlated with the galaxy's stellar mass as well as halo mass. Therefore, any radial segregation of these properties may at least partially be the result of mass segregation. Dynamical friction \citep{Chandrasekhar43} is predicted to play an important role in driving mass segregation \citep{Ostriker75,Tremaine75,White77}; The resultant drag force increases with mass resulting in massive objects being preferentially found near the centre of the group or cluster.

The existence of mass segregation in different environments is still a topic of debate. Observational studies such as \citet{Lares04} analyzed the dynamical properties of group galaxies from 2dFGRS and found significant segregation trends by examining the differences in the velocity functions of galaxies of different luminosity (and therefore stellar mass) ranges. \citet{vandenBosch08} also found stellar mass segregation trends with projected radius using data from the Sloan Digital Sky Survey Data Release 7 (SDSS-DR7) \citep{Abazajian09}. They concluded that segregation trends in colour and concentration naturally arise due to mass segregation and the correlation between stellar mass and colour/concentration. \citet{Roberts15} also found evidence for weak (stellar) mass segregation using galaxy groups in SDSS-DR7. They concluded that the mass segregation trend is strengthened by the inclusion of low mass galaxies and that the trend is weaker in higher mass groups/clusters. On the other hand, \citet{vonDerLinden10} studied cluster galaxies in SDSS and found no evidence of stellar mass segregation in their sample. \citet{Ziparo13} only found a mild segregation trend in stellar mass in the low redshift end of their sample of X-ray selected groups from the COSMOS, GOODS and ECDFS fields.

Studies of simulated galaxies have also explored mass segregation. \citet{Contini15} used dark-matter simulations along with semi-analytic models in 4 different host mass regimes ranging from $10^{13}\ h^{-1}M_{\odot}$ to greater than $5\times10^{14}\ h^{-1}M_{\odot}$. They found that within a virial radius, the mean galaxy mass decreases with halo-centric distance while between $1-2\,R_{vir}$, it increases with distance. Most recently, \citet{Bosch16} conducted an extensive study of the segregation of various properties of subhalos in the Bolshoi and Chinchilla simulations. Their sample consisted of host halos with a minimum mass of $6.7\times10^{12}h^{-1}M_{\odot}$ from the Bolshoi simulation and $7.2\times10^{12}h^{-1}M_{\odot}$ and $3.0\times10^{13}h^{-1}M_{\odot}$ for the two Chinchilla simulations. They found a weak correlation between the subhalos' present day mass and their location in the larger host halo, although they found other indicators such as the mass at infall and the amount of mass lost after infall to be more strongly correlated with radius. 

The lack of consensus among these studies is partly due to the differences in the way mass segregation is measured in observational data vs. simulated data. Observational studies generally focus on trends in stellar mass and projected radial separation from halo centres; studies of simulations use 3D separations \citep[although note that several authors do also look at projection effects and how these can alter their results, e.g.][]{Bosch16}. Simulation studies also generally consider dark matter halo masses, since the conversion from halo mass to stellar mass requires either an assumed stellar mass-halo mass relation, semi-analytical modeling or sophisticated hydrodynamical simulations. Each of these techniques can add scatter to the relation due to the additional assumptions regarding star formation and feedback. 

Understanding mass segregation requires an understanding of the assembly history of the system. Observationally, we cannot follow an individual galaxy over its entire lifetime. Instead, we observe galaxies at different epochs and infer the processes occurring over time. Another challenge is that although $\sim50\%$ of galaxies at low redshifts are associated with a group or cluster, groups become difficult to detect without extensive spectroscopy over large areas. Cosmological simulations can be used to help overcome these challenges, as we are able to identify halos at every time step and track the evolution of a single halo through its lifetime. Simulations also provide full phase-space information which could be key in finding group/cluster members.

In order to use simulations to study galaxy evolution, it is critical to first robustly identify halos and subhalos. Several different techniques have been used to identify halos in simulations. One set of early halo finders used a Friends-of-Friends (FOF) algorithm that links together particles separated by distances smaller than a linking length `b', usually specified as a fraction of the mean interparticle spacing in the simulation \citep{Davis85}. These early algorithms would occasionally link two distinct halos through a tenuous bridge of particles and were also unable to detect substructure due to the use of a single linking length \citep{Knollmann09}. Spherical overdensity (SO) algorithms did not suffer from the first problem. These algorithms identify particles around density peaks and determine a terminal radius at which the average density within the sphere is the critical density of the Universe multiplied by a factor that comes from the spherical collapse model \citep{Lacey94}. Nearly all current halo finders are descendants of one these two fundamental algorithms. 

The next step is to detect substructure within the overdensities of the distinct halos. Several algorithms that can detect subhalos now exist including Bound Density Maxima (BDM) \citep{Klypin97}, Subhalo Finder (SUBFIND) \citep{Springel01}, Amiga's Halo Finder (AHF) \citep{Knollmann09} and ROCKSTAR \citep{Behroozi13}. In this work, we examine the performance of AHF and ROCKSTAR in detecting substructure for the purpose of studying mass segregation. Previous studies have tended to only include direct `subhalos' i.e. a single level lower than the parent halo, and sometimes `subsubhalos' of the host halo in their analysis; here we investigate the consequences of such criteria. There have been past efforts in comparing the performance of various halo finders e.g. \citet{Lacey94,Cole96,White02,Knollmann09,Lukic09,Muldrew11,Behroozi13}. We compare our results to the recent work of \citet{Knebe11} with the \emph{Haloes gone MAD} project and \citet{Onions12} with the \emph{Subhaloes gone Notts} project. \citet{Knebe11} carried out an extensive comparison project, using several popular halo finders including AHF and ROCKSTAR, studying their ability to accurately reconstruct particle memberships, centres, masses, bulk velocities and dispersions etc. from mock data as well from an actual simulation. They found that both AHF and ROCKSTAR were able to recover the masses of the mock halo and subhalo to within $\sim5\%$ although only phase-space halo finders such as ROCKSTAR were able to detect subhalos at very small separations from the host halo centre. \citet{Onions12} examined the properties of subhalos embedded in a single Milky Way-like halo from the Aquarius project \citep{Springel08} using several different halo finders. They found good agreement between AHF and ROCKSTAR in terms of the subhalo mass functions and radial distributions they detected.

Comparing the mass and radial distributions of the subhalos detected by both halo finders prompts the need to select a new population of `galaxy analogues' that better corresponds to observed galaxy populations. We then use these galaxy analogues to search for possible mass segregation trends not only in the total sample, but also separating by host halo mass in order to study possible environmental effects. The paper is organized as follows: in Section \ref{sec:sim}, we provide the details of the simulation and describe the halo finding algorithms used in this study. In Section \ref{sec:hmf} we look at the mass functions of halos and subhalos detected by both halo finders at all three resolutions. Section \ref{sec:rad} examines the radial distributions of subhalos and motivates the need to select galaxy analogues. Section \ref{sec:ga} describes the selection criteria for these analogues and examines their mass and radial distributions. Finally, with this new population, we look for possible mass segregation trends and the effects of host mass and low mass `galaxies' in Section \ref{sec:massSeg}.

\section{Methods} \label{sec:sim}

\subsection{Simulation}

\begin{figure}
\centering
\includegraphics[height=0.9\textheight]{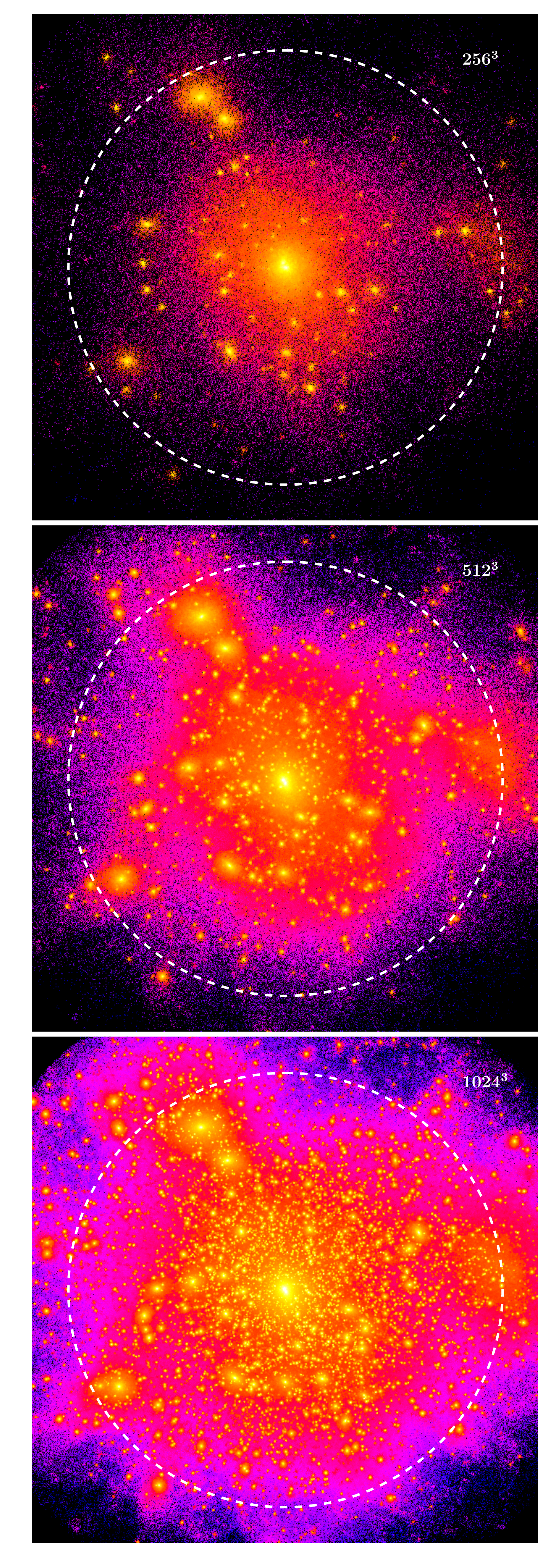}
\caption{Projected density of the most massive distinct halo at all three resolutions. The white dotted line shows virial radius $r_{vir}=2.18$ Mpc. Particles are coloured according to a smoothed density on a logarithmic scale. Centre coordinates and radius were taken from the highest-resolution run. The same large scale features are recovered at all three resolutions.} \label{fig:densityMap}
\end{figure}

We performed collisionless, cosmological N-body simulations using the Tree-SPH code ChaNGa \citep{Jetley08,Jetley10,Menon14} at three different resolutions. The simulation volume was a comoving box of length $100$ Mpc on each side. The low-, medium- and high-resolution runs contained $256^{3}$, $512^{3}$ and $1024^{3}$ particles respectively, with gravitational softening lengths of $5$ kpc, $2.5$ kpc and $1.25$ kpc, resulting in particle masses of $2.4\times10^{9}M_{\odot}$, $2.9\times10^{8}M_{\odot}$ and $3.7\times10^{7}M_{\odot}$ respectively. The softening lengths are comoving for $z>8$, physical at lower redshifts. Initial conditions (ICs) were generated using MUSIC \citep{Hahn13} assuming a flat $\Lambda$CDM cosmology with $n_{s}=0.9611$, $\sigma_{8}=0.8288$, $\Omega_{\Lambda}=0.6814$, $\Omega_{m}=0.3086$, $h=0.6777$ \citep[the cosmological parameters were obtained from][]{Planck14}. The same parameters and random number seeds were used at all three resolutions in order to ensure that we recovered the same large scale structures. Note that MUSIC requires mass and dimension values to be in $h$-inverse units whereas ChaNGa does not. We chose to work in absolute units for the entire analysis; since the halo finders also output masses and radii etc. in $h$-inverse units, we first convert all quantities to absolute units using the value of $h$ from \citet{Planck14}. Each simulation was started at a redshift of $z=100$ and evolved to $z=0$ in $1000$ linear timesteps (output every 25 timesteps). For a first look at the results, Fig. \ref{fig:densityMap} shows projected density maps of the most massive halo at all three resolutions (the centre coordinates and radius are taken from the highest-resolution run). The overall large-scale structures are recovered well at all three resolutions.

\subsection{Halo finding}
We first compare the performance of halo finders that use both spatial and velocity information vs. those that only use spatial information. The velocity information may not provide much additional help in identifying isolated halos that are spatially well separated. However, in high density environments where several halos are found in a small volume, as well as when multiple halos live within a larger host halo, halo finders which use velocity information have been shown to be better at finding distinct halos and subhalos \citep[e.g.][]{Knebe11}. We used two representative algorithms - AHF which is a spatial algorithm, and ROCKSTAR which is a phase space algorithm. The two codes employ different techniques to identify potential halos and subhalo hierarchies as described below. In both halo finders, the virial radius is defined as the radius within which the average density is given by
\begin{equation}
\bar{\rho}_{vir}(z) = \Delta_{c}(z)\rho_{c}(z) = \Delta_{m}(z)\rho_{m}(z)
\end{equation}
where $\rho_{c}$ is the critical density of the Universe and $(\rho_{m}=\Omega_{m}\rho_{c})$ is the background matter density at the given redshift. The factor $\Delta_{c}$ is calculated for a flat matter-$\Lambda$ Universe following \citet{Bryan98} as 
\begin{equation}
\Delta_{c}(z) = 18\pi^{2} + 82x - 39x^{2}
\end{equation}
where
\begin{equation}
x = \frac{\Omega_{m,0}(1+z)^{3}}{\Omega_{m,0}(1+z)^{3}+\Omega_{\Lambda}} - 1
\end{equation}
With the cosmological parameters used in this study we find $\Delta_{c}=102$ and $\Delta_{b}=332$ at $z=0$ .

\subsubsection{AMIGA's Halo Finder}
AHF identifies halos as spherical overdensities in the spatial distribution of particles in simulations. The major steps in the algorithm are as follows. (See \citet{Knollmann09} for further details.)
\begin{enumerate}
\item An Adaptive Mesh Refinement (AMR) grid is generated starting from a coarse regular grid; any cell whose particle density exceeds a specified threshold (\emph{NperRedCell=5}) is refined by splitting it into 8 equally-sized cells. The process is repeated until no cell exceeds the particle density threshold.
\item Starting at the finest refinement level, contiguous regions at each refinement level are marked as potential halos. A grid hierarchy is also built whereby each potential halo at a finer refinement level is linked to the region it resides in at the (coarser) level above it.
\item When multiple potential halos at one level live within the same region at an upper level, the potential halo with the most number of particles is assigned to be the `host'; the rest are designated `subhalos'. This process is repeated at every refinement level.
\item The `leaves' of this grid hierarchy, i.e. the halos at the finest refinement levels, are assumed to be the centres of potential halos. Particles within a given radius are assigned to these centres. For a host halo with no subhalos, this radius is the first isodensity contour where the density is lower than the required $\bar{\rho}$. For subhalos, the radius is half the distance to the host halo.
\item An iterative unbinding procedure is performed to remove particles with speeds greater than the escape velocity $\times$ a tunable factor. Unbound particles from a subhalo are considered for boundedness to the host halo. This is done for all potential halo centres to determine a final list of halos. Only halos with a minimum number of particles (\emph{for this study, NminPerHalo=20}) are kept in the final output. 
\item Further halo properties are calculated using the bound particles assigned to the halo. Note that subhalo particles are included when calculating `host' halo properties. By construction, all particles within the virial radius are bound to the halo.
\end{enumerate}

\subsubsection{ROCKSTAR}
ROCKSTAR identifies halos as overdensities in the 6D phase-space distribution of the simulation particles, using a friends-of-friends (FOF) algorithm. The major steps in the algorithm are as follows. (See \citet{Behroozi13} for further details.) Note that although there are a number of tunable parameters whose values we specify below, we use the default values recommended by \citet{Behroozi13} as they have been extensively tested.
\begin{enumerate}
\item Initially, overdense regions in the spatial distribution of particles are identified using a modified fast 3D FOF algorithm with a single large linking length. This is done only in order to break up the simulation into independent units that can be further analyzed in parallel.
\item Within a single overdense region (`parent group'), phase-space overdensities are identified using a 6D FOF method with a distance metric defined as
\begin{equation}
d(p_1,p_2) = \sqrt{\frac{\left|\vec{x}_{1}-\vec{x}_{2}\right|^{2}}{\sigma_{x}^{2}} + \frac{\left|\vec{v}_{1}-\vec{v}_{2}\right|^{2}}{\sigma_{v}^{2}}} \label{eq:distMetric}
\end{equation}
where $\sigma_{x}$ and $\sigma_{v}$ are the dispersions in the particle positions and velocities for the region. The linking length is chosen such that a specific fraction of particles (\emph{FOF\_FRACTION=0.7}) are linked to at least one other particle.
\item The process is repeated at each level generating a hierarchy of subgroups; a tighter linking length is chosen at each level, corresponding to a higher overdensity. It is terminated when a group reaches a specified minimum number of particles (\emph{MIN\_HALO\_PARTICLES=10}).
\item The FOF groups at the finest refinement level become `seed halos'. If a parent group contains a single seed halo, all particles in the parent group are assigned to the seed. If multiple seeds exist in a single parent group, particles are assigned to the closest seed halo in phase-space using a modified distance metric.
\item Halo-subhalo relations are determined by treating the `seed halos' as single particles and calculating a modified distance metric to all halos with larger numbers of assigned particles. The halo in question is then assigned to be the subhalo of the nearest larger halo in phase space.
\item An unbinding procedure is carried out to remove unbound particles. Halo properties are then calculated using only bound particles. Only halos with a minimum number of assigned particles are included in the final output (\emph{for this study, MIN\_HALO\_OUTPUT\_SIZE=20}). Also not included in the final output are any halos whose fraction of unbound particles exceeds a maximum threshold (\emph{UNBOUND\_THRESHOLD=0.5})
\end{enumerate}

\section{Mass functions}	\label{sec:hmf}

\begin{table}
\caption{Numbers of distinct halos, subhalos and galaxy analogues identified by both halo finders in the high-resolution simulation, in various mass ranges. The numbers in brackets are the average value for a single parent halo. Since analogues are identified out to $3\,r_{vir}$ while subhalos only extend to $1\,r_{vir}$, we also provide numbers of galaxy analogues within $1\,r_{vir}$} \label{tab:haloCounts}
\begin{tabular}{|l|r|r|}
\hline
Subset	& ROCKSTAR	& AHF	\\
\hline
\multicolumn{3}{|c|}{Distinct halos}	\\
\hline
$M_{vir} \geq M_{complete} (=10^{9}M_{\odot})$ & 787,374 & 810,082 \\ 
$M_{vir} \geq 10^{12.5}M_{\odot}$ & 606 & 604 \\ 
$10^{12.5} \leq M_{vir} < 10^{13}M_{\odot}$ & 411 & 409 \\ 
$10^{13} \leq M_{vir} < 10^{14}M_{\odot}$ & 178 & 178 \\ 
$10^{14} \leq M_{vir} < 10^{15}M_{\odot}$ & 17 & 17 \\ 

\hline
\multicolumn{3}{c}{Subhalos with $M_{vir} \geq 10^{10}M_{\odot}$ and $M_{parent} \geq 10^{12.5}M_{\odot}$}	\\
\hline
Total & 10,145 (17) & 11,535 (20) \\ 
$10^{12.5} \leq M_{par} < 10^{13}M_{\odot}$ & 2,501 (7) & 2,659 (7) \\ 
$10^{13} \leq M_{par} < 10^{14}M_{\odot}$ & 4,382 (25) & 4,849 (28) \\ 
$10^{14} \leq M_{par} < 10^{15}M_{\odot}$ & 3,262 (192) & 4,027 (237) \\ 

\hline
\multicolumn{3}{c}{Galaxy analogues with $M_{parent} \geq 10^{12.5}M_{\odot}$}	\\
\hline
Total & 43,864 (36) & 44,070 (36) \\ 
$10^{12.5} \leq M_{par} < 10^{13}M_{\odot}$ & 10,349 (26) & 10,311 (26) \\ 
$10^{13} \leq M_{par} < 10^{14}M_{\odot}$ & 18,925 (107) & 18,753 (106) \\ 
$10^{14} \leq M_{par} < 10^{15}M_{\odot}$ & 14,590 (859) & 15,006 (883) \\ 

\hline
\multicolumn{3}{c}{Galaxy analogues with $M_{parent} \geq 10^{12.5}M_{\odot}$ and $r<r_{vir}$}	\\
\hline
Total & 13,726 (23) & 14,213 (24) \\ 
$10^{12.5} \leq M_{par} < 10^{13}M_{\odot}$ & 2,910 (8) & 3,374 (9) \\ 
$10^{13} \leq M_{par} < 10^{14}M_{\odot}$ & 5,644 (32) & 6,152 (35) \\ 
$10^{14} \leq M_{par} < 10^{15}M_{\odot}$ & 5,172 (305) & 4,687 (276) \\ 

\hline
\end{tabular}
\end{table}

\begin{figure*}
\includegraphics[width=\linewidth]{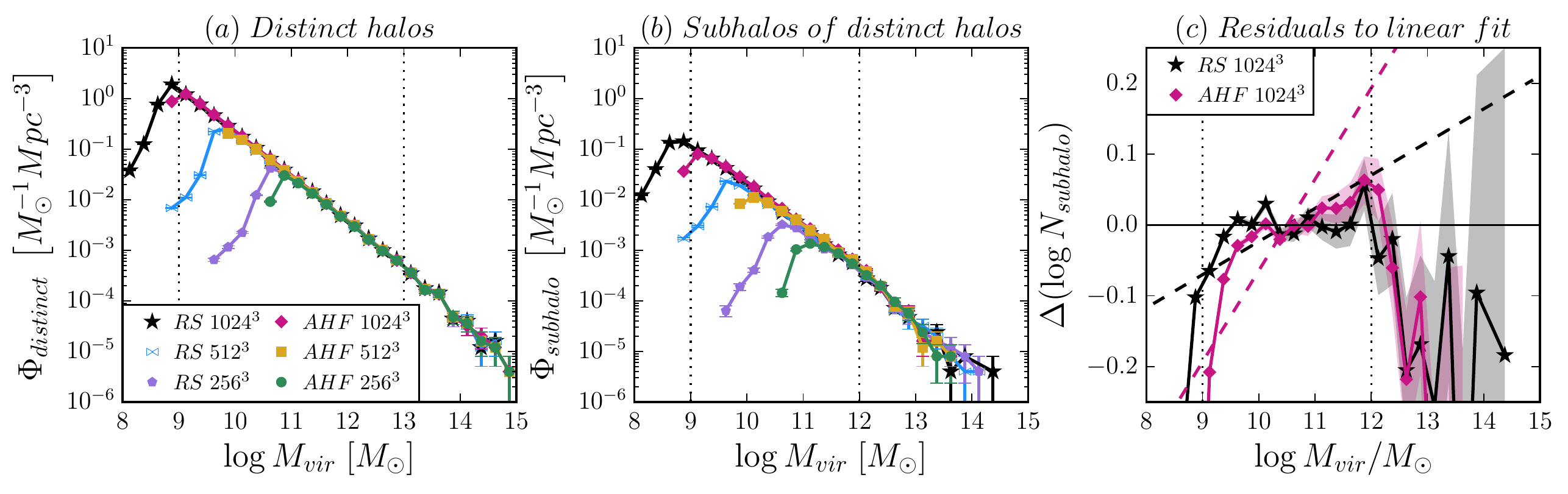}
\caption{\emph{(a)}: Mass functions (MFs) of distinct halos identified by both halo finders and at all three resolutions. $\mathbf{\Phi=dn/d\log{M_{vir}}}$ is the number density of halos per unit $\log{(mass)}$. \emph{(b)}: MFs of subhalos; colours are identical to (a). The uncertainties in both these plots are calculated as Poisson errors. \emph{(c)} Residuals of subhalo MFs and corresponding distinct halo MFs to show differences in their slopes. We fit a power-law model of the form $\left[\mathbf{\log{\Phi} = a\log{M_{vir}}+b}\right]$ to the MFs within the mass ranges indicated by the vertical dotted lines in (a) and (b) (these limits are explained in the text). The fits from (a) are then normalized at $M=10^{10.5}M_{\odot}$ to match the corresponding value in (b). The data points show the residuals between the subhalo MFs and the renormalized distinct halo MF fits. The dashed lines show the difference between the subhalo MF fits and the renormalized distinct halo MF fits. For both halo finders, the subhalo MF appears to be shallower than the corresponding distinct MF, although the difference for ROCKSTAR is smaller than for AHF.} \label{fig:hmf}
\end{figure*}

\begin{figure*}
\includegraphics[width=\linewidth]{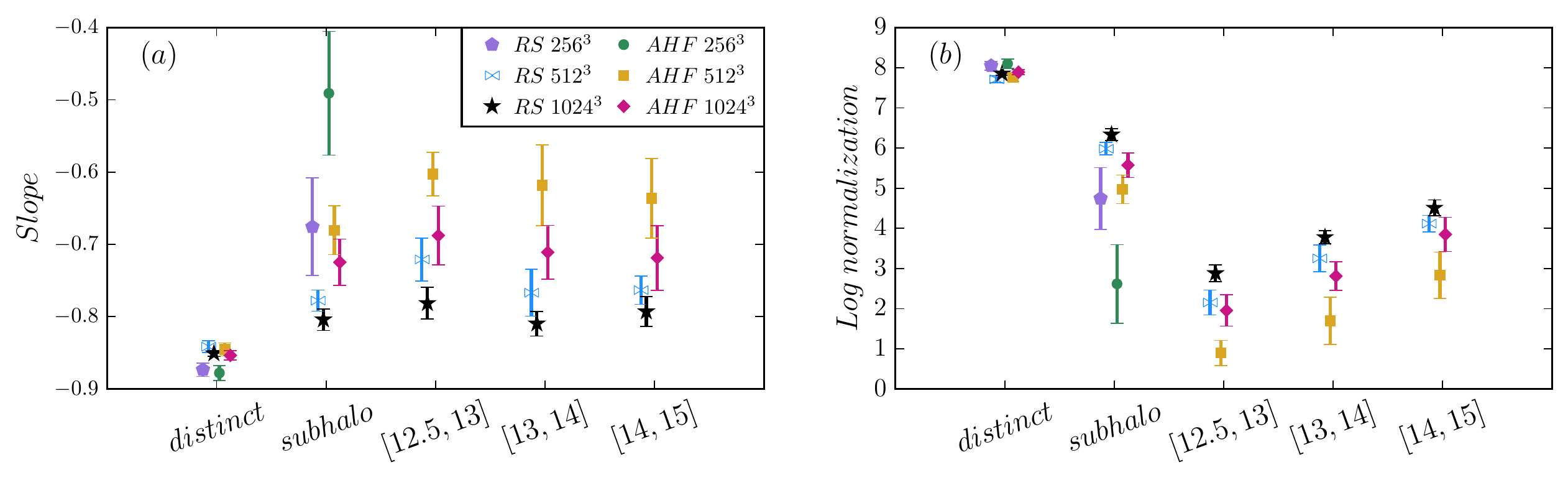}
\caption{\emph{(a)}: Log slopes and \emph{(b)}: log-normalizations from power-law fits to the mass functions shown in Figs. \ref{fig:hmf} and \ref{fig:hmfBin}. The various samples shown here are the distinct halos, the complete sample of subhalos and the three subhalos samples in bins of host mass $\log{M_{parent}}=[12.5,13]$, $[13,14]$ and $[14,15]$ respectively. Arbitrary offsets were applied in the x-direction for clarity. We only use data points within a specific range in $\log{M_{vir}}$ - the lower limit is set by requiring a minimum of 25 particles in each halo, which corresponds to 11,10 and 9 for the low-, mid- and high-resolutions runs respectively; the upper limit is set by requiring a maximum relative error of 10\%, which corresponds to 13 for the distinct halos, 12 for the total sample of subhalos and 11.25 for the binned subhalos samples.} \label{fig:powerFit}
\end{figure*}

We first compare the overall mass distributions of the host halos and their satellites as identified by the two halo finders through their mass functions (MFs). Throughout this paper, the set of `distinct' halos are those that do not have a parent i.e. they are at the top level of the halo hierarchy. Of these, only distinct halos with $M_{vir}\geq 10^{12.5}M_{\odot}$ are taken to be part of the `parent' halo set. Based on the $M_h-M_*$ relations in \citet{Hudson15}, this corresponds to a stellar mass of roughly $M_*=10^{11}M{\odot}$; any halos more massive than this are expected to contain more than a single galaxy and would therefore qualify as host halos. The \emph{direct} subhalos of these parent halos, i.e. only one level down in the subhalo hierarchy, whose centres lie within one virial radius from the parent halo centre are considered to be the `subhalo' population. For reference, Table \ref{tab:haloCounts} provides the numbers of distinct halos and subhalos detected by both halo finders in the highest-resolution simulation within various mass ranges.

\subsection{Distinct halos} \label{sec:iso}
Fig. \ref{fig:hmf}(a) shows the MFs of the distinct halos detected at all 3 resolutions and by both halo finders. The errors shown are Poisson errors. Qualitatively, the two halo finders produce consistent results down to the completeness limits of $M_{vir}=(10^{11},10^{10},10^{9})M_{\odot}$ for the low-, mid- and high-resolution runs respectively. These completeness limits are set by requiring that each halo is resolved by a minimum of 25 particles. The ROCKSTAR MFs extend to lower masses than AHF. This is because in AHF, the user specifies the minimum \emph{bound} particles a halo must have to be included in the final output (20 in this study); in ROCKSTAR the minimum threshold specifies how many particles have been uniquely assigned to the FOF group, but due to the aspherical nature of these groups, a significant number of these particles can lie outside the virial radius, which means the virial mass for the halos can be lower than the mass of 20 particles. However, these differences are at masses lower than the completeness limit and therefore do not impact our analysis.

The MFs appear to be a single power law down to the completeness limit with no flattening evident at the low mass end. At the high mass end, there does appear to be some evidence of the exponential drop off expected for a Schechter profile, but we are limited by low number statistics and cannot probe this region with any certainty. For a quantitative comparison, we instead fit a single power law between the completeness limit for the particular resolution and $M_{vir}=10^{13}M_{\odot}$. The upper mass limit is set by requiring a maximum relative uncertainty of 10\%, which is equivalent to having at least 100 halos in each bin. The results of the power law fits for both halo finders and all 3 resolutions are shown in Figure \ref{fig:powerFit}. (These values are also provided in Table \ref{tab:powerFit} in the Appendix.)

Considering each resolution separately, we find that the two halo finders produce identical MFs for distinct halos both in terms of slope and normalization. The MFs for the mid- and high-resolution runs are nearly identical as well; their slopes agree within $1\sigma$ and normalizations within $2\sigma$. For the low-resolution run however, we find a systematically steeper slopes as well as a higher normalization; this is partly due to less substructure created in the simulations. 

\citet{Knebe11} find similar agreement between AHF and ROCKSTAR using a mock data set. They provide three different systems with known masses, centres and velocity offsets - (i) a single halo, (ii) a halo with an embedded subhalo and (iii) a halo with an embedded subhalo and as well as a subsubhalo - as input and compare the results from several different halo finders. The recovered masses for the isolated halos from AHF and ROCKSTAR are within $\sim5\%$ of the input values as well as each other, although note that these masses were not the values returned by the halo finders themselves. Instead, \citet{Knebe11} use the locations of the halos and the particles belonging to them to calculate halo properties using a single code in order to eliminate any differences in the way these properties are calculated by the halo finders.

There are a number of high resolution simulations of large cosmological volumes that we can compare our results to, such as the Bolshoi simulation \citep{Klypin11} and the Illustris project (specifically the Illustris-1-Dark simulations) \citep{Vogelsberger14}. While the MFs for these simulations were both fit by an empirically derived formula from \citet{Sheth02}, the simulations were large enough to have over 10 times the number of halos that we have in our simulation. They therefore had much better statistics at the high mass end where we expect a steepening of the power law. Since \citet{Klypin11} and \citet{Vogelsberger14} fit a modified Press-Schechter functional form to their MFs, they do not provide equivalent values for their slopes. However, we have estimate the MF slopes for both Bolshoi and Illustris within a mass range of $(10^{9}-10^{13})\ h^{-1}M_{\odot}$ and found them to be in agreement with our results.

\subsection{Subhalos}
We next look at the MFs of the subhalos, shown in Fig. \ref{fig:hmf}(b). We measure the power-law slopes of the MFs over a mass range defined by the completeness limit at the low-mass end and $10^{12}M_{\odot}$ at the high-mass end. The upper limit here is lower than the one used for distinct halos in keeping with the requirement of a maximum relative error of 10\%. The results of the fits are provided in Fig. \ref{fig:powerFit} as well as Table \ref{tab:powerFit} in the Appendix. 

We find that the low-mass completeness limits set for the distinct halos remain applicable for the subhalos as well. The shapes of the MFs are qualitatively the same as the ones for the distinct halos - they obey power-law relations upto the upper mass limit used for the model fit. The turn-over at the high mass end does appear to be at a lower mass than in the case of the distinct halos. At each resolution, the AHF MFs have slopes that are systematically shallower and consistently lower normalizations as compared to the ROCKSTAR results indicating that AHF detects slightly fewer low mass subhalos. Setting aside the low-resolution run, for a single halo finder, the two higher resolution runs have statistically identical MFs, although the best fit values for the slopes are consistently shallower and the best fit normalizations are lower in the mid-resolution run than in the high-resolution run.

In order to compare the slopes of the subhalo MFs and distinct halo MFs, we first remove the obvious differences stemming from their normalizations. We normalize the power-law fit to the distinct halo MF in Fig. \ref{fig:hmf}(a) at $M=10^{10.5}M_{\odot}$ to match the corresponding value of the subhalo MF. The differences between the subhalo MFs and the renormalized distinct halo MF fits for the high-resolution run are plotted in Fig. \ref{fig:hmf}(c). The dashed lines show the difference between the subhalo MF fits and the renormalized distinct halo MF fits. Fig. \ref{fig:hmf}(c) shows that the subhalo MFs are shallower than the distinct halo MFs, although the difference is more pronounced in the case of AHF than in ROCKSTAR. This may be expected as more massive subhalos have higher density peaks that can be distinguished more easily from the background halo density profile, making them easier to detect than lower mass subhalos.

\subsection{Dependence on parent mass}
Since the parent halos cover a large mass range, it is possible that the results found above could vary with parent halo mass. We look at the same MFs separated into bins of parent halo mass $M_{parent}=10^{12.5}-10^{13}M_{\odot},$ $10^{13}-10^{14}M_{\odot}$ and $10^{14}-10^{15}M_{\odot}$. The numbers of subhalos in each bin for the high-resolution simulation are given in Table \ref{tab:haloCounts}. The bin widths were chosen to have comparable numbers of subhalos in each bin as well as representing different environments. The lowest parent-mass bin might represent a poor group or a large galaxy with a population of dwarfs around it; the intermediate bin might represent a group of a few to a few tens of bright galaxies; the highest parent-mass bin would represent a cluster of galaxies.  As before, we fit power-law models within a mass range defined by the completeness limits and $10^{11.25}M_{\odot}$. The results of the fits are shown in Fig. \ref{fig:powerFit}. (There were too few points within this mass range in the low-resolution run to calculate a good fit, which is why it is excluded from these fits.)

As with the total subhalo sample, the MF slopes at both resolutions agree within $1\sigma$ for both halo finders separately, although the best fit slopes for the mid-resolution run are systematically shallower than the ones for the high-resolution run. The AHF slopes are systematically shallower and normalizations lower than ROCKSTAR. The normalizations increase steadily with parent halo mass which is to be expected as more massive host halos should contain more substructure. Crucially, the slopes are identical in all three environments when controlling for resolution and halo finder.  Thus, we find no significant effect of environment on the subhalo MFs. For completeness, we provide the MFs in different parent-mass bins and comparisons to the distinct halo MF in Fig. \ref{fig:hmfBin} in the Appendix.

\section{Radial distributions of subhalos}	\label{sec:rad}

\begin{figure*}
\includegraphics[width=\linewidth]{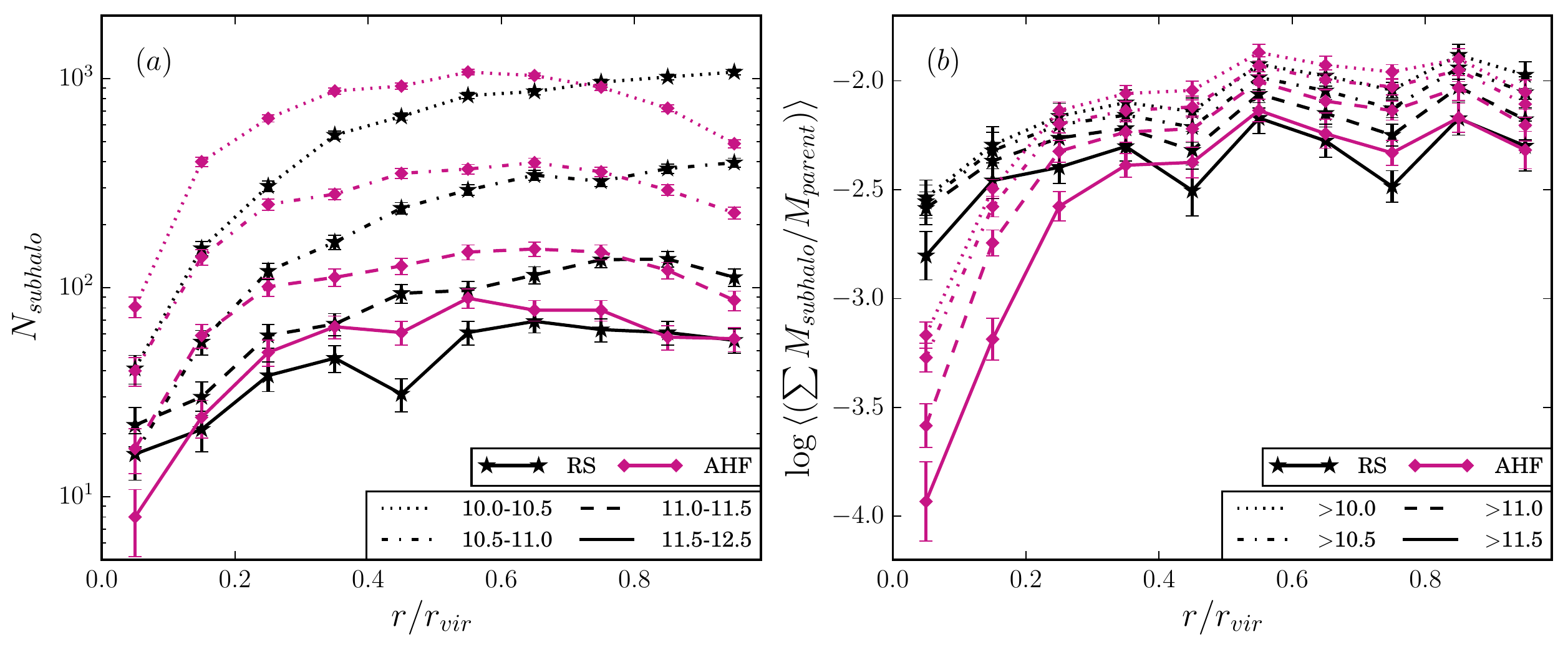}
\caption{\emph{(a)}: Stacked numbers of subhalos in bins of radial separation from the centre of the parent halo. The different linestyles correspond to different bins of $\log{M_{subhalo}/M_{\odot}}$ while the colours correspond to the two halo finders used as in previous figures. AHF and ROCKSTAR produce different radial trends, with AHF generally finding more subhalos within $\sim0.7\,r_{vir}$ at all subhalo masses except in the innermost radial bin. The effect is largest in the lower subhalo mass bins. \emph{(b)}: Fractional subhalo mass vs. radial distance from parent halo centre, averaged by the total number of parent halos. The linestyles again represent bins of $\log{M_{subhalo}/M_{\odot}}$; here they all have the same upper limit of $M_{subhalo}=10^{12.5}M_{\odot}$ while the lower limit is gradually changed to show the contribution of lower mass subhalos. Low mass halos contribute very little to the total subhalo mass within the host, which is why in the inner $\sim0.5\,r_{vir}$, ROCKSTAR assigns more mass to subhalos that AHF does even though this mass comes from fewer subhalos.} \label{fig:hmfRadCount}
\end{figure*}

While the mass functions of subhalos detected by both halo finders appear to be consistent, environmental effects on the subhalos will depend on their radial position within the parent halo. For the remainder of the paper, we focus only on the highest resolution run. Fig. \ref{fig:hmfRadCount}(a) shows the stacked radial distribution (3D) of the subhalos. The sample is broken up into bins of $log{(M_{subhalo})}$, shown as different linestyles in the figure. The distance of each subhalo from the centre of the parent halo is normalized by the virial radius of the parent halo so that equal weighting is given to both high mass and low mass parent halos. From Fig. \ref{fig:hmfRadCount}(a), it is clear that although the subhalo populations selected by both halo finders appeared similar in their mass distributions, their radial distributions are quite different. Despite not using velocity information, AHF surprisingly finds many more subhalos than ROCKSTAR within $0.7\,r_{vir}$. The only place this trend is reversed is in the innermost radial bin in the two higher subhalo mass bins. 

Fig. \ref{fig:hmfRadCount}(b) shows the radial profile of the total fractional mass in subhalos (i.e. $M_{subhalo}/M_{parent}$), normalized by the total number of parent halos. The different linestyles here represent bins of $\log{M_{subhalo}}$ that have the same upper limit of $10^{12.5}M_{\odot}$ while the lower limit is gradually changed in order to show the contribution of lower mass subhalos. Fig. \ref{fig:hmfRadCount}(b) shows that within $\sim0.5\,r_{vir}$ ROCKSTAR assigns more mass to the subhalos than AHF does (nearly an order of magnitude higher in the first radial bin). Even though ROCKSTAR detects fewer subhalos in the inner regions, they appear to represent a larger \emph{fraction} of the mass within their parent halos. This is a result of the fact that although the low mass subhalos dominate the number of subhalos, they do not make up much of the total mass in subhalos. \citet{Knebe11} found that close to the halo centre, only phase-space halo finders like ROCKSTAR could detect subhalos at all, though they also tended to over- or underestimate the mass in the subhalo. This may explain why ROCKSTAR assigns more mass to subhalos near the centre in our results, although it contradicts the larger numbers of subhalos found by AHF near the centre.

We also explored the dependence of these results on the host mass and found similar results as those of the total sample. AHF found more subhalos than ROCKSTAR in the inner regions though the differences are more pronounced as the mass of the parent halos increases. The fractional mass in subhalos is higher for ROCKSTAR than AHF in the lowest parent-mass bin but the trend appears to reverse in the highest parent-mass bin; these results are dominated by the most massive subhalos. Radial profiles of the numbers of subhalos and their fractional mass in bins of parent halo mass are shown in Fig. \ref{fig:hmfRadCountBin} in the Appendix. 

Fig. \ref{fig:hmfRadCount} shows that the two halo finders detect significantly different radial distributions of subhalos. There are two possible interpretations of these results - either the two halo finders detect different satellite populations altogether or the designation of `subhalo' selects different subsets of the true satellite population. The latter would mean that the `subhalo' designation is insufficient to select a sample of satellites that would match an observed galaxy population. In the next section, we therefore explore different selection criteria to construct a more representative sample of satellites.

\section{Galaxy analogues}	\label{sec:ga}

\begin{figure*}
\includegraphics[width=\linewidth]{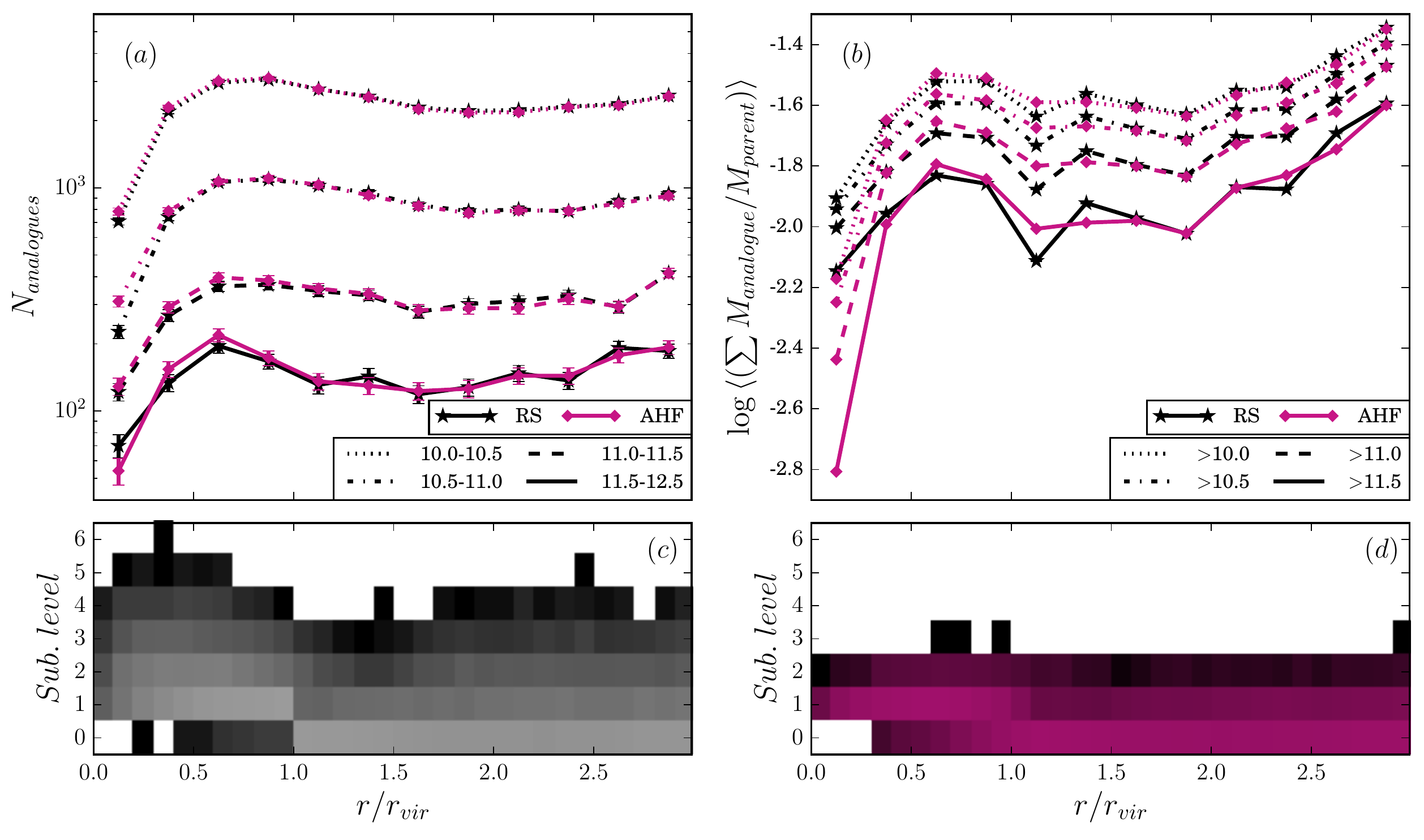}
\caption{\emph{(a)}: Stacked numbers of analogues in bins of radial separation from parent halo centre. The linestyles correspond to different bins of $\log{M_{analogue}/M_{\odot}}$ while the colours correspond to the two halo finders as in Fig. \ref{fig:hmfRadCount}. \emph{(b)}: Fractional analogue mass vs. radial separation from parent halo center, averaged by the total number of parent halos. The selection criteria for analogues ensure that we select the same population not only in terms of mass, but also in terms of radial distribution, although within $\sim0.5\,r_{vir}$, ROCKSTAR still assigns more mass to subhalos than AHF does. \emph{(c) \& (d)}: 2D histograms of radial distance between analogues and the centre of the parent halo on the x-axis and subhalo level on the y-axis. (Distinct halos are level 0, subhalos level 1, subsubhalos level 2 etc.). For (c) \& (d), brighter colours indicate higher numbers with black corresponding to a single halo and the brightest gray in (c) and the brighest pink in (d) corresponding to $1675$ and $1680$ analogues respectively. The colour maps are logarithmic and continuous with a range of $1-10^{4}$ analogues. While the same population can be selected consistently from either halo catalogue, the analogues can be at very different levels in the subhalo hierarchy. ROCKSTAR identifies many more levels of subhalos than AHF.} \label{fig:hmfGACount}
\end{figure*}

\begin{figure*}
\includegraphics[width=\linewidth]{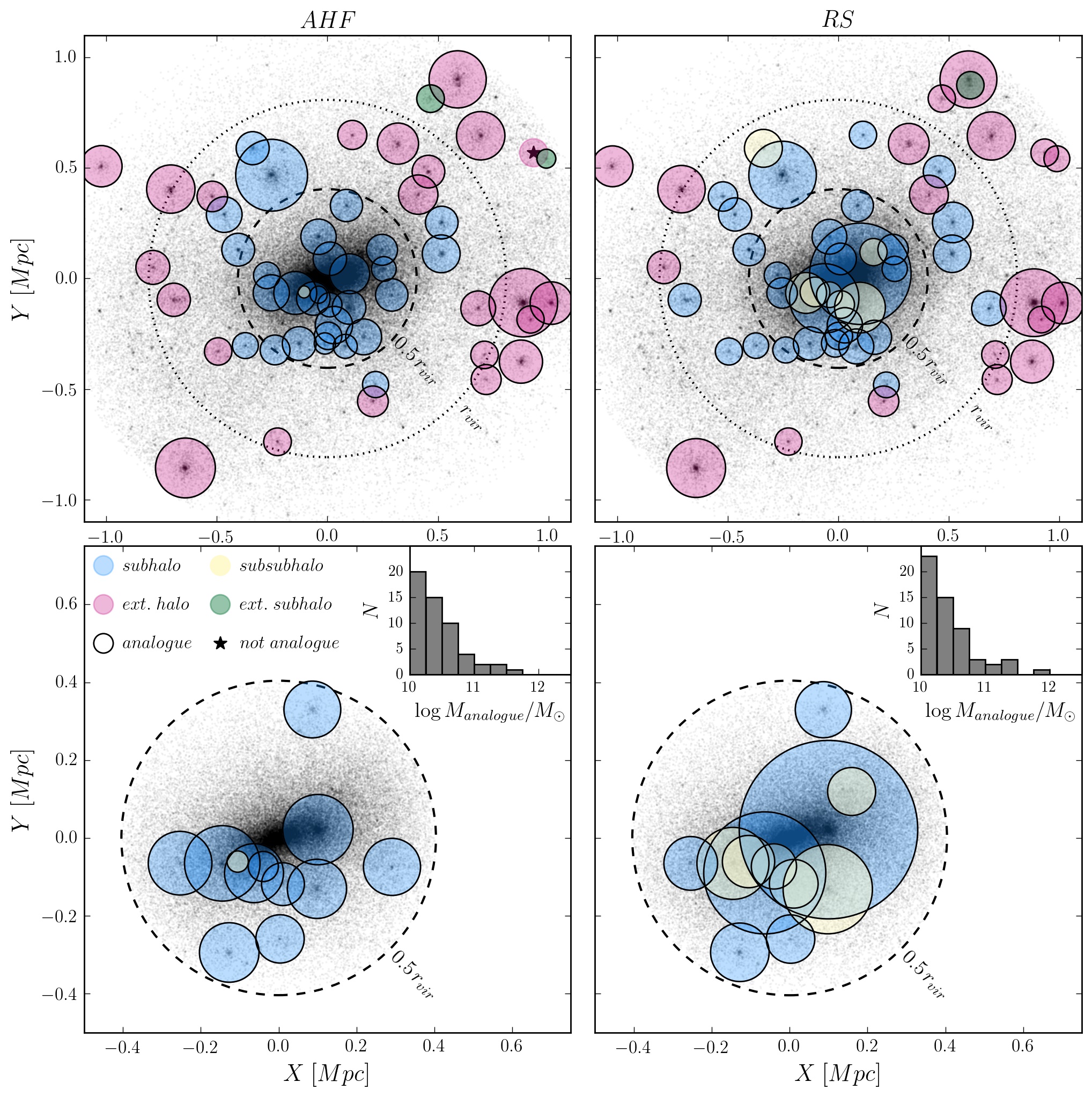}
\caption{Comparing (sub)subhalos and analogues identified by AHF (\emph{Left}) and ROCKSTAR (\emph{Right}) for an example system with mass $M_{vir}=2.9\times10^{13}M_{\odot}$. \emph{Top}: Particles within $1.5\,r_{vir}$ from the parent halo centre are plotted as black points (we only plot every 5\textsuperscript{th} particle for clarity). $r_{vir}$ for the host halos is shown by the black dotted circle, $0.5\,r_{vir}$ by the black dashed circle. Coloured circles show the satellites around the halo - subhalos of the main parent halo are shown in blue, subsubhalos in yellow; distinct halos around the main parent halo within $1.5\,r_{vir}$ are shown in red while their subhalos are shown in green. The radii of the circles match their virial radii. We only show (sub)subhalos with $M_{vir}>10^{10}M_{\odot}$ here. \emph{All} analogues identified for the system within $1.5\,r_{vir}$ are shown by the solid black circles. Nearly all of the satellites were selected as analogues. The one exception in the case of AHF is shown by the star symbol. \emph{Bottom}: Zooming in on the inner $0.5\,r_{vir}$. The insets also show the MF of the total analogue population within $1.5\,r_{vir}$. While the sets of subhalos and subsubhalos from both halo finders are different, the analogue populations are more consistent, although within $0.5\,r_{vir}$ even analogue populations look different. The figures show that a selection based solely on `subhalo' designations would result in several analogues being missed.} \label{fig:AHFvsRS}
\end{figure*}

It is clear that the subhalo populations identified by the two halo finders show differences in their radial distributions, yet the global mass functions are in agreement. One of the reasons for this may be that by only selecting subhalos, we are neglecting the deeper levels in the subhalo hierarchy, i.e. several of the subhalos have subsubhalos of their own that could host galaxies. Additionally, in observational studies, analysis is carried out on visible galaxies whereas the simulated dark matter subhalos could host 0, 1 or multiple galaxies. Therefore, it is important to select the right halos in order to be consistent with observational results. To do this we employ a simple method of selecting such a population of `galaxy analogues' (hereafter referred to as just `analogues'). 

Our main selection criterion for analogues is mass; a halo is only eligible to be a galaxy analogue if its mass lies within the range of $M_{vir}=[10^{10},10^{12.5}M_{\odot}]$, which would roughly correspond to a stellar mass of $M_{*}=[10^{8.5},10^{11}]M_{\odot}$ based on the $M_{h}-M_{*}$ relations in \citet{Hudson15}. However, a simple mass cut is insufficient since there may be cases where both a halo and its subhalos meet this criterion, in which case we need to examine whether both are likely to host galaxies. We assign levels to each halo based on its position in the subhalo hierarchy - distinct halos are designated level 0, subhalos level 1, subsubhalos level 2 etc. We then start with a distinct halo and work through its subhalo hierarchy. Any candidate for an analogue is put through the following selection criteria. The `halo' here can be at any level in the hierarchy. 
\begin{enumerate}
\item If $M_{vir} < 10^{10}M_{\odot}$, then the halo and its subsequent branches in the hierarchy are eliminated.
\item If $M_{vir} \geq 10^{12.5}M_{\odot}$, then the halo itself is ignored, but each of its subhalos is considered as an analogue candidate and put through these same selection criteria.
\item If the halo has a mass $10^{10}<M_{vir}<10^{12.5}M_{\odot}$ and either has no subhalos or all of its subhalos have masses $M_{vir} < 10^{10}M_{\odot}$, then the halo is included as an analogue while its subsequent branches are eliminated.
\item If the halo has a mass $10^{10}<M_{vir}<10^{12.5}M_{\odot}$ and at least one of its subhalos has a mass $M_{vir} > 10^{10}M_{\odot}$, then we first look at the excess mass it contains after subtracting off the masses of \emph{all} of its subhalos (not just the ones that meet the mass criteria). If this excess mass is also within the valid range, then the halo is included as an analogue and each of its subhalos is also considered as an analogue candidate.
\end{enumerate}
Effectively, if we have a system in which a halo (at any level) and one or more of its subhalos both meet the mass criterion, then the final step means that in a limited number of cases, we keep both the halo as well as its subhalos. Once the analogue population has been identified, we select analogues within $3\,r_{vir}$ of every parent halo centre. The numbers of analogues identified by both halo finders in various parent-halo mass ranges are provided in Table \ref{tab:haloCounts}.

The distribution of the final analogue population is shown in Fig. \ref{fig:hmfGACount}. The two top panels are identical to Fig. \ref{fig:hmfRadCount}, only now showing analogues instead of subhalos. The distributions of the total numbers of analogues from both halo finders are in much better agreement using this selection strategy. The bottom two panels are 2D histograms of the analogues with radial distance from the parent halo centre on the x-axis and subhalo level on the y-axis. It is clear from Fig. \ref{fig:hmfGACount} (c) and (d) that the analogues selected represent much deeper levels of the subhalo hierarchy in the ROCKSTAR catalogue than in the AHF one. Crucially, some of the analogues selected here are at much deeper levels that what would be included by a `subhalo' or `subsubhalo' designation. We also examined these trends in the three different parent-mass bins and the results are included in Fig. \ref{fig:hmfGACountBin} in the Appendix. We find that the results from both halo finders also agree well within all three environments, except possibly in the innermost radial bins in the lowest parent mass bin. Again, the ROCKSTAR analogues represent deeper levels of the subhalo hierarchy as compared to the AHF ones in all three environments. 

In Fig. \ref{fig:AHFvsRS} we show an example system with mass $M_{vir}=2.9\times 10^{13}M_{\odot}$, to explicitly compare the subhalo hierarchies. In the top panels, we plot the particles within $1.5\,r_{vir}$. Ignoring any (sub)subhalos with $M_{vir}<10^{10}M_{\odot}$, we plot the subhalos in blue and subsubhalos in yellow. We also show in red other distinct halos within $1.5\,r_{vir}$ of the parent halo centre and their subhalos in green. We refer to all four sets collectively as `satellites'. We plot \emph{all} analogues within $1.5\,r_{vir}$ as black open circles, so that any satellite outlined in black is also an analogue. For both AHF and ROCKSTAR, all but one of the satellites were also analogues. The one satellite that was not included in the AHF case is designated by the star symbol. In the bottom panels we focus on the inner $0.5\,r_{vir}$. The insets also show the MF of the analogues. Fig. \ref{fig:AHFvsRS} shows how several subhalos that are designated as `subhalos' by AHF are considered `subsubhalos' by ROCKSTAR while also being `galaxy analogues'. A selection based on subhalo designations alone would therefore miss significant portions of the analogue population. 

Figs. \ref{fig:hmfGACount} and \ref{fig:AHFvsRS} confirm that while the sets of subhalos and subsubhalos are different between the two halo finders, the analogue populations are consistent with one another outside $\sim0.5\,r_{vir}$. AHF appears more likely to break up nearby overdensities into distinct subhalos at a shallower level, while ROCKSTAR is more likely to group them into a bigger subhalo and then assign them as subsubhalos at a deeper level. Therefore, we find that even without using velocity information, a halo finder like AHF can detect most of the halos of interest outside $\sim0.5\,r_{vir}$ as well as a halo finder like ROCKSTAR can. In the inner regions however, phase-space information is crucial in being able to separate substructure from the main host halo as was found by previous studies such as \citet{Knebe11}.

\section{Mass segregation}	\label{sec:massSeg}

\begin{figure*}
\includegraphics[width=\linewidth]{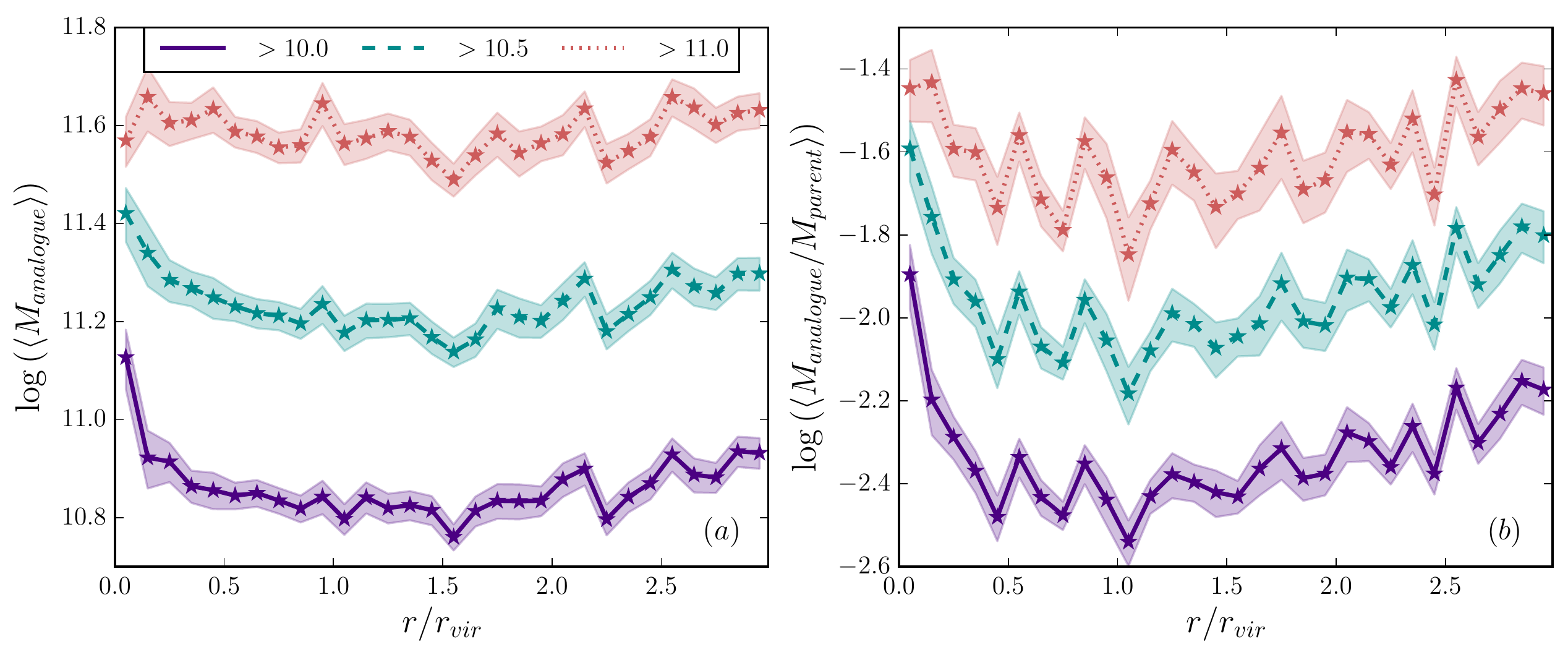}
\caption{Radial trends in average analogue mass (\emph{Left}) and average fractional analogue mass (\emph{Right}) in mass bins of $\log{(M_{analogue}/M_{\odot})}$, as shown by the different colours and linestyles. The bins have the same upper limit of $M_{analogue}=10^{12.5}M_{\odot}$ while the lower limit is gradually changed to show the effect of including lower mass halos in detecting mass segregation trends. Errors shown are standard errors on the mean. Within $0.5\,r_{vir}$ we find a weak trend with average mass decreasing with radius. There is also a milder trend outside $r_{vir}$ with average mass increasing with radius. The trends are sensitive to the lower mass limit applied to the sample - they are stronger when low mass analogues are included (purple solid line and green dashed line). The trends are also more prominent when looking at fractional mass, again only within $0.5\,r_{vir}$.} \label{fig:GAMassSeg}
\end{figure*}

\begin{figure*}
\includegraphics[width=\linewidth]{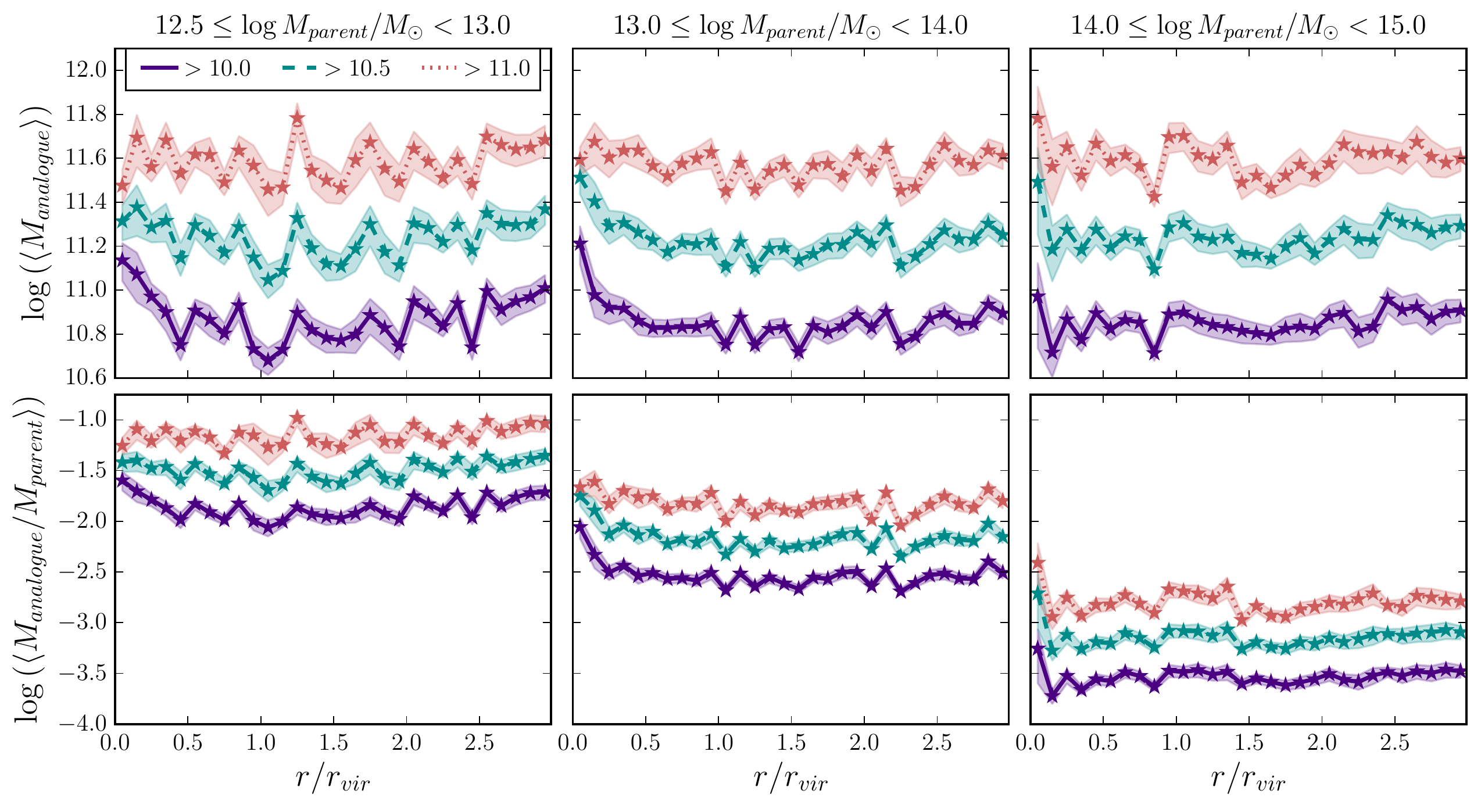}
\caption{Radial trends in average analogue mass (\emph{Top}) and average fractional analogue mass (\emph{Bottom}) as in Fig. \ref{fig:GAMassSeg}, but separated by environment. As with the total sample, we see a weak trend of average mass decreasing with radius within $0.5\,r_{vir}$, though only in the low- and intermediate-parent mass systems and only when we include lower mass analogues down to a mass of at least $10^{10.5}M_{\odot}$. Most importantly, the trends get weaker with increasing parent halo mass.} \label{fig:GAMassSegBin}
\end{figure*}

\begin{table*}
\caption{Results of linear fits to mass segregation trends from the ROCKSTAR results including analogues of all masses (purple solid lines in Figs. \ref{fig:GAMassSeg} and \ref{fig:GAMassSegBin}). For the average mass profile, the fit is of the form: $\mathbf{\log{(M_{analogue})} = \left[a_1\left(r/r_{vir}\right)+b_1\right]}$. For the average fractional mass profile, the fit is of the form $\mathbf{\log{(M_{analogue}/M_{parent})} = \left[a_2\left(r/r_{vir}\right)+b_2\right]}$. All uncertainties are standard errors on regression coefficients. Non-zero slopes are shown in bold.} \label{tab:massSegFit}
\begin{tabular}{lcccc}
\hline
Subset	&	$a_1$	&	$b_1$	&	$a_2$	&	$b_2$	\\
\hline
\multicolumn{5}{c}{$r/r_{vir}<0.5$}	\\
\hline
All analogues & $\mathbf{-0.54\pm 0.18}$ & $11.07\pm 0.06$ & $\mathbf{-1.28\pm 0.22}$ & $-1.93\pm 0.07$ \\ 
$10^{12.5} \leq M_{par} < 10^{13}M_{\odot}$ & $\mathbf{-0.96\pm 0.08}$ & $11.20\pm 0.03$ & $\mathbf{-0.95\pm 0.04}$ & $-1.55\pm 0.01$ \\ 
$10^{13} \leq M_{par} < 10^{14}M_{\odot}$ & $\mathbf{-0.71\pm 0.22}$ & $11.16\pm 0.07$ & $\mathbf{-0.96\pm 0.38}$ & $-2.14\pm 0.12$ \\ 
$10^{14} \leq M_{par} < 10^{15}M_{\odot}$ & $+0.18\pm 0.36$ & $10.77\pm 0.12$ & $+0.17\pm 0.43$ & $-3.67\pm 0.15$ \\ 

\hline
\multicolumn{5}{c}{$0.5<r/r_{vir}<1$}	\\
\hline
All analogues & $\mathbf{-0.05\pm 0.04}$ & $10.87\pm 0.03$ & $-0.12\pm 0.25$ & $-2.32\pm 0.19$ \\ 
$10^{12.5} \leq M_{par} < 10^{13}M_{\odot}$ & $\mathbf{-0.27\pm 0.24}$ & $11.05\pm 0.18$ & $-0.20\pm 0.32$ & $-1.76\pm 0.24$ \\ 
$10^{13} \leq M_{par} < 10^{14}M_{\odot}$ & $\mathbf{+0.04\pm 0.02}$ & $10.80\pm 0.01$ & $-0.04\pm 0.13$ & $-2.52\pm 0.10$ \\ 
$10^{14} \leq M_{par} < 10^{15}M_{\odot}$ & $-0.16\pm 0.30$ & $10.93\pm 0.23$ & $-0.00\pm 0.25$ & $-3.55\pm 0.19$ \\ 

\hline
\multicolumn{5}{c}{$1<r/r_{vir}<3$}	\\
\hline
All analogues & $\mathbf{+0.06\pm 0.01}$ & $10.72\pm 0.03$ & $\mathbf{+0.14\pm 0.02}$ & $-2.62\pm 0.04$ \\ 
$10^{12.5} \leq M_{par} < 10^{13}M_{\odot}$ & $\mathbf{+0.12\pm 0.03}$ & $10.61\pm 0.06$ & $\mathbf{+0.13\pm 0.03}$ & $-2.14\pm 0.06$ \\ 
$10^{13} \leq M_{par} < 10^{14}M_{\odot}$ & $\mathbf{+0.05\pm 0.02}$ & $10.72\pm 0.04$ & $\mathbf{+0.05\pm 0.03}$ & $-2.69\pm 0.06$ \\ 
$10^{14} \leq M_{par} < 10^{15}M_{\odot}$ & $\mathbf{+0.04\pm 0.02}$ & $10.77\pm 0.03$ & $\mathbf{+0.03\pm 0.02}$ & $-3.59\pm 0.04$ \\ 

\hline
\end{tabular}
\end{table*}

With the analogue population identified, we are able to examine whether we see any mass segregation trends in the host halos. Given that phase-space information appears to be important in identifying analogues in the inner regions of parent halos, we focus on the ROCKSTAR results here. Fig. \ref{fig:GAMassSeg}(a) shows the radial profile of average analogue mass. The different colours and linestyles represent bins of $\log{M_{analogue}}$ where the upper limit is always 12.5 (the mass cut off for selecting analogues) while the lower limit is varied to explore the effect of including low mass halos. Fig. \ref{fig:GAMassSegBin}(b) shows the average radial profile of analogue mass as a fraction of the mass of its host halo. As seen in Fig. \ref{fig:GAMassSeg}(a), the mass profile is remarkably flat over most of the $3\,r_{vir}$ considered here and any significant mass segregation trend is confined to within $\sim0.5\,r_{vir}$. 

Linear models of the form $\mathbf{\log{(M_{analogue})} = a_1\left(r/r_{vir}\right)+b_1}$ and $\mathbf{\log{(M_{analogue}/M_{parent})} = a_2\left(r/r_{vir}\right)+b_2}$ were fit to the radial profiles from $(0-0.5)\,r_{vir}$, $(0.5-1)\,r_{vir}$ and $(1-3)\,r_{vir}$ separately. The results of the fits are given in Table \ref{tab:massSegFit} (we focus on the fits for the total mass range of analogues, which corresponds to the solid purple lines in Fig. \ref{fig:GAMassSeg}). The average mass decreases with distance in the inner $0.5\,r_{vir}$, rises with distance beyond $r_{vir}$ and is nearly constant between $(0.5-1)\,r_{vir}$ with a minimum at $\sim r_{vir}$. We find a slope $a_1$ of $-0.5\pm0.2$ within $0.5\,r_{vir}$, $-0.05\pm0.04$ between $(0.5-1)\,r_{vir}$ and $+0.06\pm0.01$ between $(1-3)\,r_{vir}$. The trends are stronger when we consider fractional mass instead of absolute mass as seen in Fig. \ref{fig:GAMassSeg}(b). The corresponding slopes $a_2$ are $-1.3\pm0.2$ within $0.5\,r_{vir}$, $-0.1\pm0.2$ between $(0.5-1)\,r_{vir}$ and $+0.14\pm0.02$ between $(1-3)\,r_{vir}$.

Another effect evident in Fig. \ref{fig:GAMassSeg}(a) is that the radial trends in mean absolute mass seen at small radii are stronger when we include low mass analogues in our analysis. If we consider only high mass analogues, the results are consistent with having no trend with radius. Thus, the higher mass substructure does not appear to have a preferred position within the parent halo. Any segregation trend is instead due to the low mass substructure. Note that since we are looking at the average mass, these results are not due to having a larger volume at large radii but due to intrinsic variations of the analogue population with radial distance. In observational studies, due to detection constraints, low mass galaxies are often not included in order to have a luminosity- or mass-complete sample. Our results suggest that this can have a significant impact on whether mass segregation is detected. 

We also examine these mass segregation trends separating by parent halo mass to explore any environmental dependence. Fig. \ref{fig:GAMassSegBin} shows the same radial trends in average mass (top panels) and average fractional mass (bottom panels), now separated by environment. Results of linear fits to the profiles for the complete mass range of analogues (purple solid lines) are given in Table \ref{tab:massSegFit}. Outside $0.5\,r_{vir}$, the results are qualitatively in agreement with what we find for the total population - outside $r_{vir}$ there is a mild increase in average (fractional) mass with distance; between $(0.5-1)\,r_{vir}$, the average fractional mass profiles are consistent with having no slope while we find mild slopes in the average absolute mass in the two lower-parent-mass bins. Within $0.5\,r_{vir}$ however, we do find significant differences based on environment. In the highest parent-mass bin, both the average absolute mass and fractional mass profiles are consistent with being flat, whereas we find significant segregation in the two lower parent-mass bins. In all cases, regardless of whether we consider absolute mass or fractional mass and within all three radial regions, the trends are strongest in the least massive systems and weakest in the most massive ones.

\section{Discussion}	\label{sec:disc}

\subsection{Detecting substructure}
In this work, we have examined the ability of AHF and ROCKSTAR to detect substructure within groups and clusters in dark matter simulations. The resulting mass functions of distinct halos they find are consistent with each other, while the subhalo mass functions show significant differences. The largest differences however are seen in the radial distributions of the subhalos. Counter-intuitively, AHF detects many more `subhalos' within the mass range of interest than ROCKSTAR. This shows the sensitivity of these halo finders to various numerical choices. However, as we have shown, outside $0.5\,r_{vir}$, both halo finders are capable of detecting consistent populations of substructure as long as care is taken in selecting `galaxy analogue' populations rather than `subhalo' populations. This is due to the differences in the subhalo hierarchies built by each halo finder whereby the same galaxy analogue could be at different levels within the hierarchy and crucially, deeper than a `subhalo' or even `subsubhalo'.

Note that the halos detected by ROCKSTAR are not necessarily spherical due to the nature of a FOF algorithm. While this is less of an issue for distinct halos, it does become important for subhalos embedded in a dense environment. Regions that are spatially distant may be connected into one large structure in phase space which is why the `subhalo' designation was insufficient in our study, since the halos of interest to us were embedded at deeper levels in the subhalo hierarchy. Studies that select subhalo populations solely based on this designation would therefore miss a significant portion of the true `galaxy' population and instead, select larger numbers of massive halos relative to the low mass ones.

The selection criteria used to identify galaxy analogues resulted in a small number of cases ($\sim5\%$) where both a halo and its subhalos (at any level in the hierarchy) were both included in the analogue population. Since subhalo particles within the virial radius of the parent halo are included by both halo finders when calculating halo properties, this may result in part of the subhalo mass being included twice. Although we do not account for this in the results presented in previous sections, we repeated the same analysis after subtracting the mass in subhalos in the small number of cases where this was relevant and only using the remaining mass. Since the volumes enclosed by the virial radii of the subhalos can partially overlap each other, this was not an ideal solution; however it was a lower limit on the mass of these halos, whereas our previous results represent an upper limit, and the two results are not significantly different.

\subsection{Mass segregation}

\begin{figure}
\includegraphics[width=\linewidth]{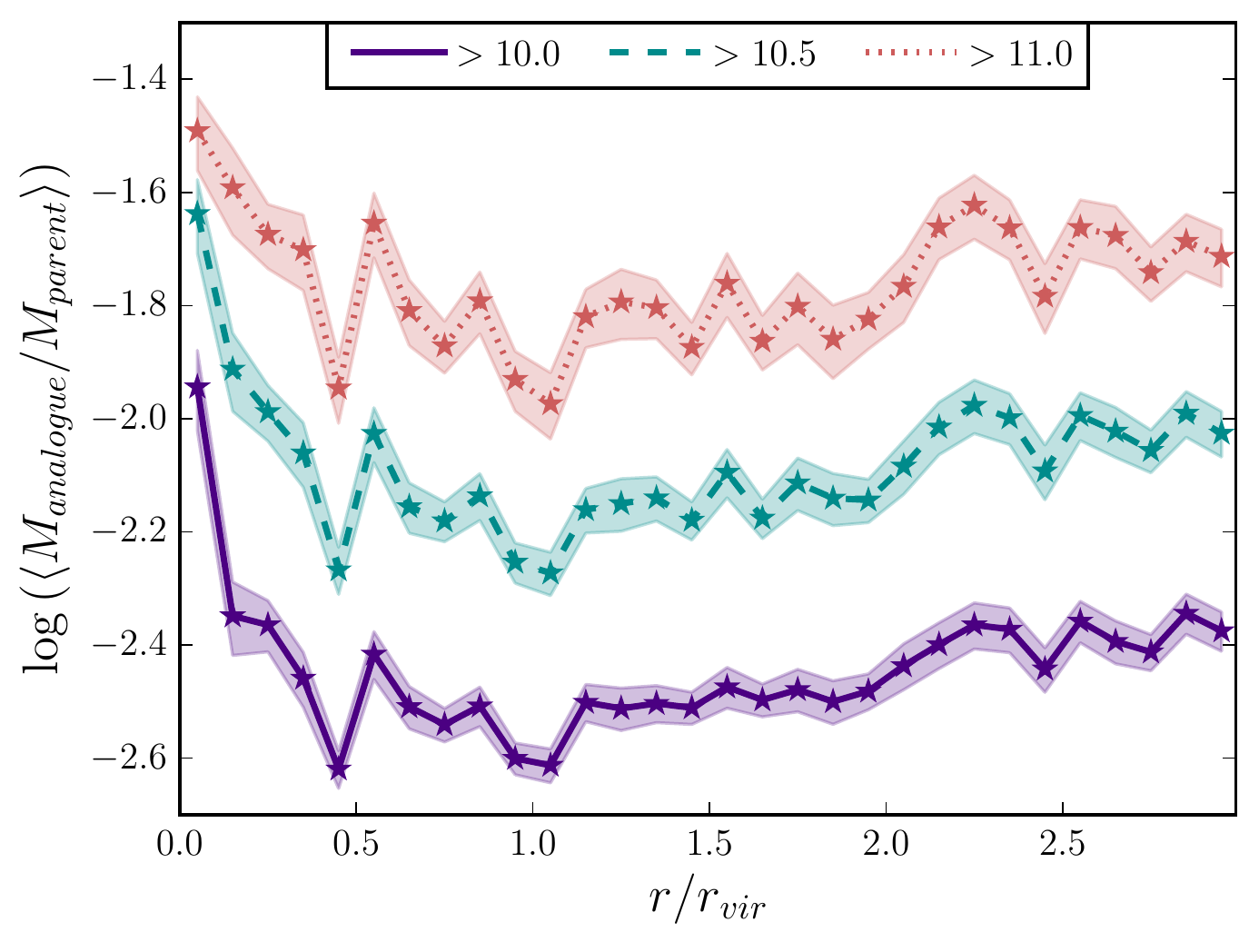}
\caption{Radial trends in average fractional analogue mass as in Fig. \ref{fig:GAMassSeg}(b) but while excluding the most massive analogues (with mass $10^{12}-10^{12.5}M_{\odot}$) to reduce stochastic scatter and show mass segregation more clearly.} \label{fig:GAMassSegUp12}
\end{figure}

\begin{figure}
\includegraphics[width=\linewidth]{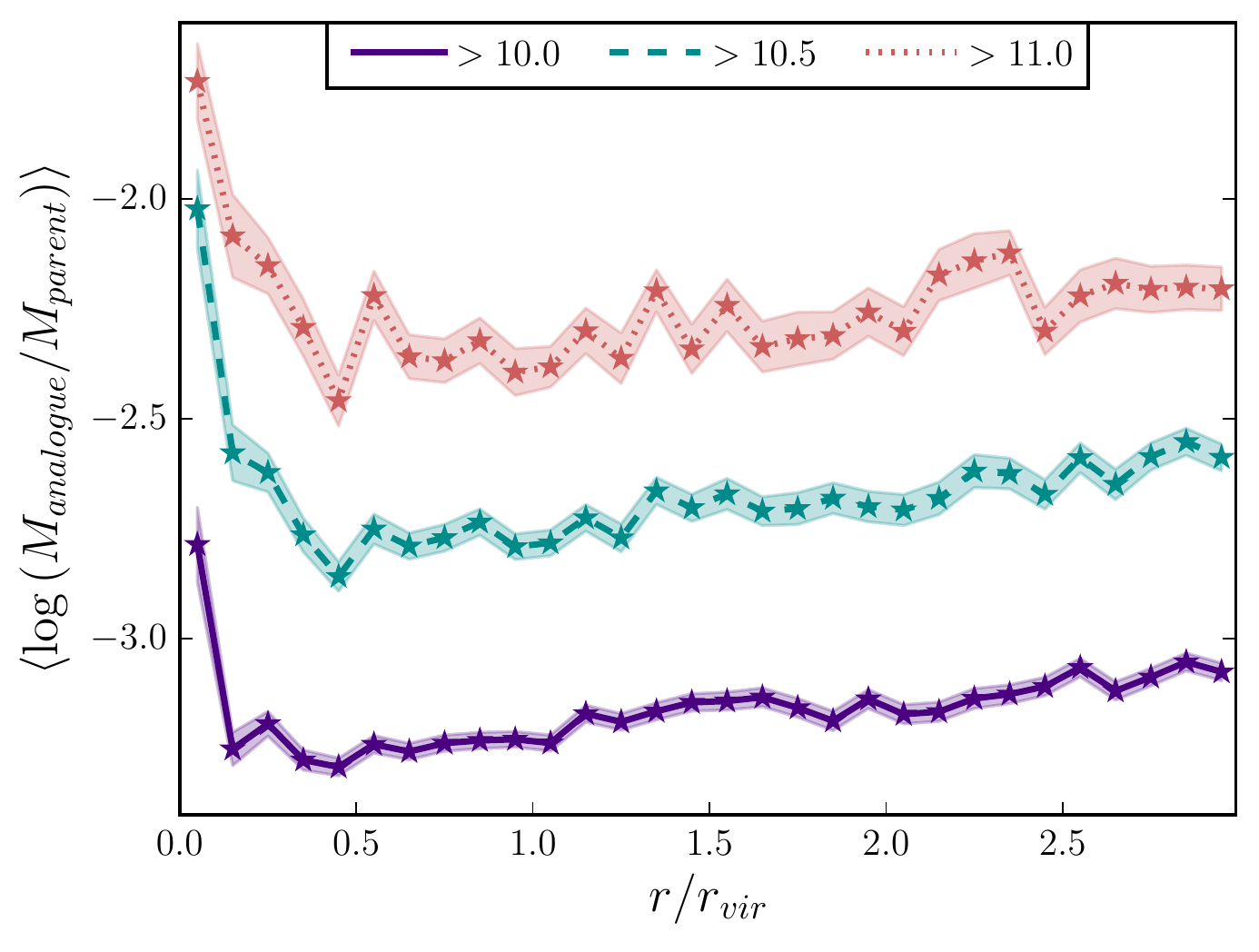}
\caption{Radial trends in average fractional analogue mass as in Fig. \ref{fig:GAMassSeg}(b) but averaging over $\log(fractional\ mass)$ rather than $(fractional\ mass)$ in order to give equal weighting to low mass analogues.}  \label{fig:GAMassSegAvgLog}
\end{figure}

We used the analogue population to examine mass segregation trends in dense environments. Firstly, in the average mass profiles shown in Fig. \ref{fig:GAMassSeg}(a), we find statistically significant trends within $0.5\,r_{vir}$ (with average mass decreasing with distance) and beyond $r_{vir}$ (with average mass rising with distance) for the total population, but only when we include low mass analogues. As seen in Fig. \ref{fig:GAMassSeg}(a) (pink dotted lines) massive analogues do not appear to preferentially live near the centres of their host halos. Separating by parent-mass, we find no mass segregation in the most massive systems, which correspond to galaxy clusters, while in less massive systems we find mild to moderate segregation. The strength of the segregation signal appears to be anti-correlated with the mass of the parent halo.

In the average mass profiles for the total population shown in Fig. \ref{fig:GAMassSeg}(b), the segregation trends are stronger and present in all three ranges of analogue mass under consideration. It is important to reconcile this with the trends in the lower panels of Fig. \ref{fig:GAMassSegBin} which appear to be shallower in comparison, especially for higher mass analogues (pink dotted lines). Each line in Fig. \ref{fig:GAMassSeg}(b) is a weighted sum of the corresponding lines in the three lower panels of Fig. \ref{fig:GAMassSegBin} with the weights equal to the fraction of analogues contributed by each parent-mass bin to each radial bin. We found that for all three ranges of analogue mass, beyond $0.5\,r_{vir}$ the largest contribution comes from the intermediate parent-mass bin. However, the contribution from the lowest parent-mass bin rises steadily with radius while the highest parent-mass bin decreases with radius. Since the absolute mass range of analogues represents a different fraction of the parent mass in each of the three systems, even if the trends in each parent mass bin had been flat, the average fractional mass profile would show a gradual increase with radius as the lowest parent-mass bin would be weighted more and more heavily. The opposite is true within $0.5\,r_{vir}$ where the contribution from the lowest parent-mass bin rises sharply towards the centre and that from the highest parent-mass bin drops significantly; this effect is partly due to fewer analogues overall near the centres of the parent halos. Any trends in the fractional mass profile in the lowest parent-mass bin are therefore amplified in the profile for the total sample. 

The radial profiles shown in Fig. \ref{fig:GAMassSeg}(b) are also somewhat noisy making it difficult to discern any trends. We have investigated the source of this noise and found that it is due to the inclusion of the most massive analogues, with $M_{analogue}>10^{12}M_{\odot}$; due to low numbers, they appear stochastically in some radial bins. Since they are included in calculating all three lines shown in the figure, they have a large impact on the resultant mass profile. If we exclude these high mass analogues, the segregation trends are stronger and less noisy as shown in Fig. \ref{fig:GAMassSegUp12}. Additionally, in both Figs. \ref{fig:GAMassSeg}(b) and \ref{fig:GAMassSegUp12}, we calculate the average mass of analogues which naturally gives higher weighting to the high mass analogues. An alternative approach is to average over $\log{(M_{analogue}/M_{parent})}$ thereby weighting high- and low- mass analogues equally. The segregation trends are stronger using this method as shown in Fig. \ref{fig:GAMassSegAvgLog}.

\citet{Bosch16} found a mild \emph{positive} correlation between fractional mass $M/M_{host}$ and 3D separation $r/r_{vir}$ within the virial radius using a Spearman rank-order correlation coefficient. However this indicator of segregation is only meaningful when the correlation between the two quantities is monotonic which was not the case. Accounting for this, they found a weak trend of decreasing mass with increasing radius which is highly sensitive to sample selection. They did however find much stronger and more robust trends when using the mass at accretion $M_{acc}/M_{host}$, the peak mass of the subhalos $M_{peak}/M_{host}$ or the mass lost by the subhalo after accretion, quantified by $M/M_{acc}$ as the segregation property. They concluded that the mild segregation in present-day mass is due to a combination of the inside-out assembly of the host halos resulting in more massive halos accreted at earlier times being found at smaller radii, dynamical friction causing massive halos to migrate towards the halo centre, as well as mass loss due to tidal stripping that acts to negate this segregation. Our results for the total analogue sample are roughly consistent with those of \citet{Bosch16}, although when we separate our sample into different parent mass bins we see no trend in the most massive parent halos. Their subhalo population was defined as halos whose centres were within $r_{vir}$ of a larger halo, which may more closely resemble our analogue population than a simple `direct subhalo' definition. Note that they excluded subhalos where $M/M_{host}<10^{-3}$ as opposed to a lower absolute mass limit and had additional constraints on $M_{acc}/M_{host}$. As the segregation trends in present day mass were highly sensitive to sample selection, these constraints almost certainly influenced their results.

\citet{Contini15} found a clear mass segregation signal of dark matter subhalos out to $1R_{vir}$ and then an upturn in average mass out to $2R_{vir}$;  the upturn was quite sharp in their `large groups' and `large clusters' samples. They considered subhalos with \emph{stellar} masses greater than $10^{10}M_{\odot}$, which is significantly more massive than the lower mass limit used in this study and the range of host masses they consider only overlap with our two higher parent-mass bins. Our results do support a mild positive mass segregation trend with halo-centric radius beyond a virial radius; however the upturn is not as sharp as the one found by \citet{Contini15}. They found that the segregation trends became weaker with larger host masses which is consistent with our findings.

Our galaxy analogue results agree well with recent observational work by \citet{Roberts15} who looked at the segregation of galaxy stellar mass using observational data from the SDSS-DR7 group catalogue. Their sample consisted of host halos with masses $M=(10^{13}-10^{15})M_{\odot}$, which is similar to our sample, and 4 different lower mass limits in $M_{*}$. They found the weakest trends in the most massive groups and the strongest in the least massive ones. They also concluded that the trend gets stronger with the inclusion of lower mass galaxies, consistent with what we see in these dark-matter-only simulations.

There are several factors which could play a role in establishing the trends we detect. Firstly, dynamical friction is one of the main candidates for driving mass segregation by preferentially moving massive analogues towards the centres of their host halos. The efficiency of dynamical friction is expected to increase with analogue mass, but decrease with parent halo mass \citep{Chandrasekhar43,Boylan-Kolchin08}. Secondly, mergers in group and cluster halos can lead to the creation of more massive analogues which are then subject to dynamical friction. Thirdly, it is important to keep in mind that analogues which were accreted earlier (when the host halos were smaller and less massive) will preferentially be located at small radii. If more massive analogues are accreted at high redshifts, this scenario can also produce mass segregation. Finally, in addition to mass growth, analogues can also lose mass due to tidal stripping which would result in dynamical friction being less efficient. The mass of an analogue and its position within its host is therefore a complicated combination of accretion time, host halo mass and accretion history, dynamical friction, mergers and stripping.

We have shown that the trends in mass segregation are strongest with the inclusion of low mass analogues. In fact, there is very little segregation in the most massive analogues, as seen in Figs. \ref{fig:GAMassSeg} and \ref{fig:GAMassSegBin}, implying that they do not have a preferred location within their parent halo. This lack of segregation in massive analogues suggests either that dynamical friction is not the dominant effect at work or that the segregation due to dynamical friction is balanced by the accretion of more massive objects at late times (and therefore at large radii). The segregation trends we observe in low mass analogues, where dynamical friction is expected to be weak, may be due to tidal stripping preferentially destroying low mass objects. However, it is unclear whether tidal stripping can be more efficient for smaller analogues than more massive ones. The low-mass analogue results are consistent with dynamical friction, although they may also be the result of coordinated infall due to the late accretion of smaller groups. The reduction in mass segregation with increasing parent halo mass is consistent with dynamical friction predictions, but the lack of segregation in the most massive analogues suggest that dynamical friction is not the dominant factor in the trends we observe.

It is important to note that we use 3D radial separations throughout this study whereas most observational studies use projected separations. \citet{Bosch16} found that any segregation trends are weaker when using projected separations, which may be the reason for the weaker trends found by observational studies \citep[e.g.][]{vonDerLinden10,Ziparo13}. Additionally, we look at segregation in present day mass, whereas both \citet{Bosch16} and \citet{Contini15} also examined mass at accretion which appears to be more strongly segregated. While the stellar mass of galaxies is likely to be correlated with their halo mass at accretion \citep[e.g.][]{Nagai05}, they can undergo a diverse range of processes within the group/cluster halo that will affect both their present halo mass as well as their current observable properties. The greater segregation at early times also supports a pre-processing scenario where segregation is weaker in massive systems due to the (late) accretion of massive objects, consistent with our results.

\section{Summary} \label{sec:summ}
In this paper, we explore mass segregation trends in groups and clusters using dark-matter-only simulations with two different halo finders: AHF, a 3D spherical overdensity algorithm, and ROCKSTAR, a phase-space FOF algorithm. We compare the performance of the halo finders in detecting substructure by comparing their subhalo MFs and radial distributions.
\begin{enumerate}
\item We find that the mass distributions of direct subhalos of the parent halos are consistent between the two halo finders. However, their radial distributions are significantly different - in the inner regions of the parent halos AHF finds more subhalos whereas ROCKSTAR assigns more mass to them.
\item We then identify a population `galaxy analogues' that would better correspond to observed galaxy populations. The radial distributions of these analogues are in better agreement between the two halo finders although in the inner regions (within $\sim0.5r\,_{vir}$), ROCKSTAR assigns a larger fraction of the parent mass to these analogues than AHF.
\item We find statistically significant mass segregation for the total sample of analogues; within $0.5\,r_{vir}$ the mean (fractional) mass decreases with radius while between $(0.5-1)\,r_{vir}$, it shows little variation with radius. Beyond $r_{vir}$, we find milder positive segregation with mean (fractional) mass increasing with radius.
\item Segregation trends in average absolute mass are stronger when we include low mass analogues. The trends are also stronger in the lowest parent-mass systems and weaken with increasing parent mass.
\end{enumerate}

Earlier studies have found a much stronger correlation with the mass at accretion or the peak mass of subhalos with radial separation from the centre which indicates that any mass segregation trends are strongly connected to the accretion histories of the host systems. Therefore, future work must focus on the formation and accretion histories of these analogues to disentangle the various drivers of mass segregation trends in these systems. 

\section*{Acknowledgements}
We thank the anonymous referee for their comments and insights which were very useful in improving the manuscript. We thank the National Science and Engineering Research Council of Canada for their funding. Computations were performed on the \emph{gpc} supercomputer at the SciNet HPC Consortium \citep{Loken10}. SciNet is funded by: the Canada Foundation for Innovation under the auspices of Compute Canada; the Government of Ontario; Ontario Research Fund - Research Excellence; and the University of Toronto. This work was made possible by the facilities of the Shared Hierarchical Academic Research Computing Network (SHARCNET:www.sharcnet.ca) and Compute Canada.

\bibliographystyle{mnras}
\bibliography{JoshiParkerWadsley2016}

\appendix 

\section{Mass functions of subhalos} 

\begin{figure*}
\includegraphics[width=\linewidth]{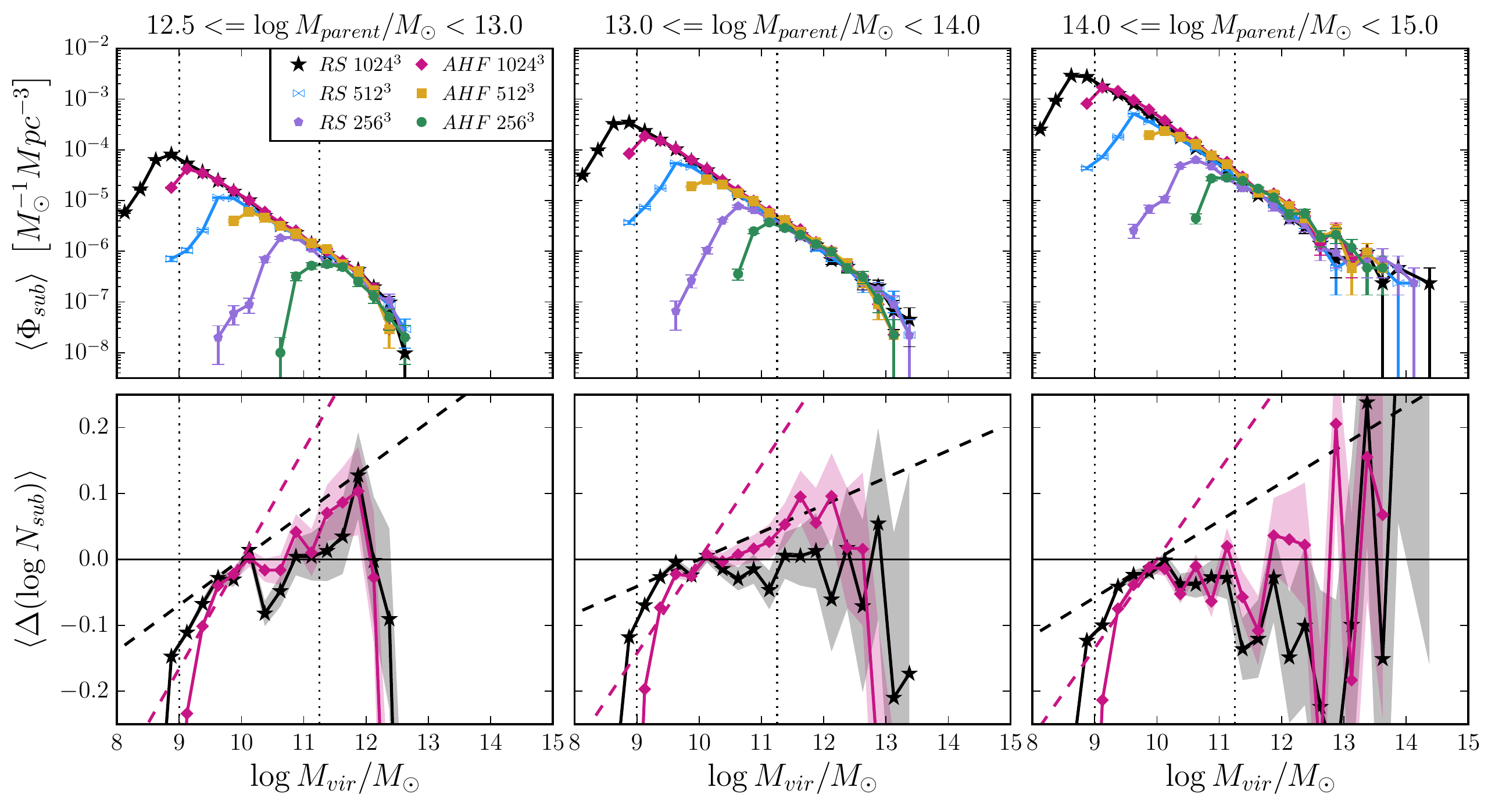}
\caption{\emph{Top}: Mass functions (MFs) and \emph{Bottom}: residuals of subhalo MFs and corresponding distinct halo MFs (as in Fig. \ref{fig:hmf} (b) \& (c) above), here separated by environment and averaged by the number of parent halos (to remove any differences caused by the different numbers of host halos in each bin). The distinct halo MFs are normalized at $M=10^{10}M_{\odot}$ here. For a single resolution and halo finder, the slopes of the MFs in the 3 bins are identical. However, the MFs for the $512^3$ run are consistently shallower than the ones for the $1024^3$ and the AHF MFs are consistently shallower than the ROCKSTAR ones. All these subhalo MFs are also consistently shallower than the corresponding distinct halo MF, although again, the differences for ROCKSTAR are smaller than for AHF.} \label{fig:hmfBin}
\end{figure*}

\begin{table*}
\caption{Results from power law fits of the form $\mathbf{\log{\Phi} = a\log{M_{vir}}+b}$ to the mass functions shown in Figs. \ref{fig:hmf} and \ref{fig:hmfBin}. We only use data points within a specific range in $\log{M_{vir}}$. The lower limit is set by requiring a minimum of 25 particles in each halo, which corresponds 11,10 and 9 for the low-, mid- and high-resolutions runs respectively. The upper limit is set by requiring a maximum relative error of 10\%, which corresponds to 13 for the distinct halos, 12 for the total sample of subhalos and 11.25 for the binned subhalos samples.} \label{tab:powerFit}
\begin{center}
\begin{tabular}{lcccccc}
\hline
&	\multicolumn{3}{c}{ROCKSTAR}	& \multicolumn{3}{c}{AHF}	\\
\hline
Subset	& $256^3$ & $512^3$ & $1024^3$ & $256^3$ & $512^3$ & $1024^3$ \\
\hline
&	\multicolumn{6}{c}{Index}	\\
\hline
Distinct halos & $-0.873\pm0.009$ & $-0.841\pm0.008$ & $-0.851\pm0.004$ & $-0.878\pm0.010$ & $-0.845\pm0.009$ & $-0.854\pm0.006$ \\ 
Subhalos & $-0.68\pm0.07$ & $-0.78\pm0.01$ & $-0.80\pm0.01$ & $-0.49\pm0.09$ & $-0.68\pm0.03$ & $-0.72\pm0.03$ \\ 
$M_{par}:\ [10^{12.5},10^{13}]M_{\odot}$ & - & $-0.72\pm0.03$ & $-0.78\pm0.02$ & - & $-0.60\pm0.03$ & $-0.69\pm0.04$ \\ 
$M_{par}:\ [10^{13},10^{14}]M_{\odot}$ & - & $-0.77\pm0.03$ & $-0.81\pm0.02$ & - & $-0.62\pm0.06$ & $-0.71\pm0.04$ \\ 
$M_{par}:\ [10^{14},10^{15}]M_{\odot}$ & - & $-0.76\pm0.02$ & $-0.79\pm0.02$ & - & $-0.64\pm0.06$ & $-0.72\pm0.04$ \\ 

\hline
&	\multicolumn{6}{c}{Normalization}	\\
\hline
Distinct halos & $8.05\pm0.11$ & $7.71\pm0.08$ & $7.85\pm0.04$ & $8.10\pm0.12$ & $7.76\pm0.09$ & $7.89\pm0.06$ \\ 
Subhalos & $4.7\pm0.8$ & $6.0\pm0.2$ & $6.3\pm0.1$ & $2.6\pm1.0$ & $5.0\pm0.4$ & $5.6\pm0.3$ \\ 
$M_{par}:\ [10^{12.5},10^{13}]M_{\odot}$ & - & $2.2\pm0.3$ & $2.9\pm0.2$ & - & $0.9\pm0.3$ & $2.0\pm0.4$ \\ 
$M_{par}:\ [10^{13},10^{14}]M_{\odot}$ & - & $3.3\pm0.3$ & $3.8\pm0.2$ & - & $1.7\pm0.6$ & $2.8\pm0.4$ \\ 
$M_{par}:\ [10^{14},10^{15}]M_{\odot}$ & - & $4.1\pm0.2$ & $4.5\pm0.2$ & - & $2.8\pm0.6$ & $3.8\pm0.4$ \\ 

\hline
\end{tabular}
\end{center}
\end{table*}

Fig. \ref{fig:hmfBin} shows subhalo MFs (\emph{Top panels}) and the residuals between the subhalo MFs and corresponding distinct halo MFs (\emph{Bottom panels}) as in Fig. \ref{fig:hmf} (b) \& (c) here separated into bins of parent halo mass. The results are qualitatively the same as those for the total subhalo sample. Power-law models were fit to these subhalo MFs within a mass range defined by the completeness limit at each resolution and $10^{11.25}M_{\odot}$ - the results are included in Fig. \ref{fig:powerFit}. 

The results of the power law fits to the distinct halo MFs, total subhalo MFs and subhalo MFs in bins of parent halo mass depicted in Fig. \ref{fig:powerFit} are also provided in Table \ref{tab:powerFit}.

\section{Radial distributions of subhalos}

\begin{figure*}
\includegraphics[width=\linewidth]{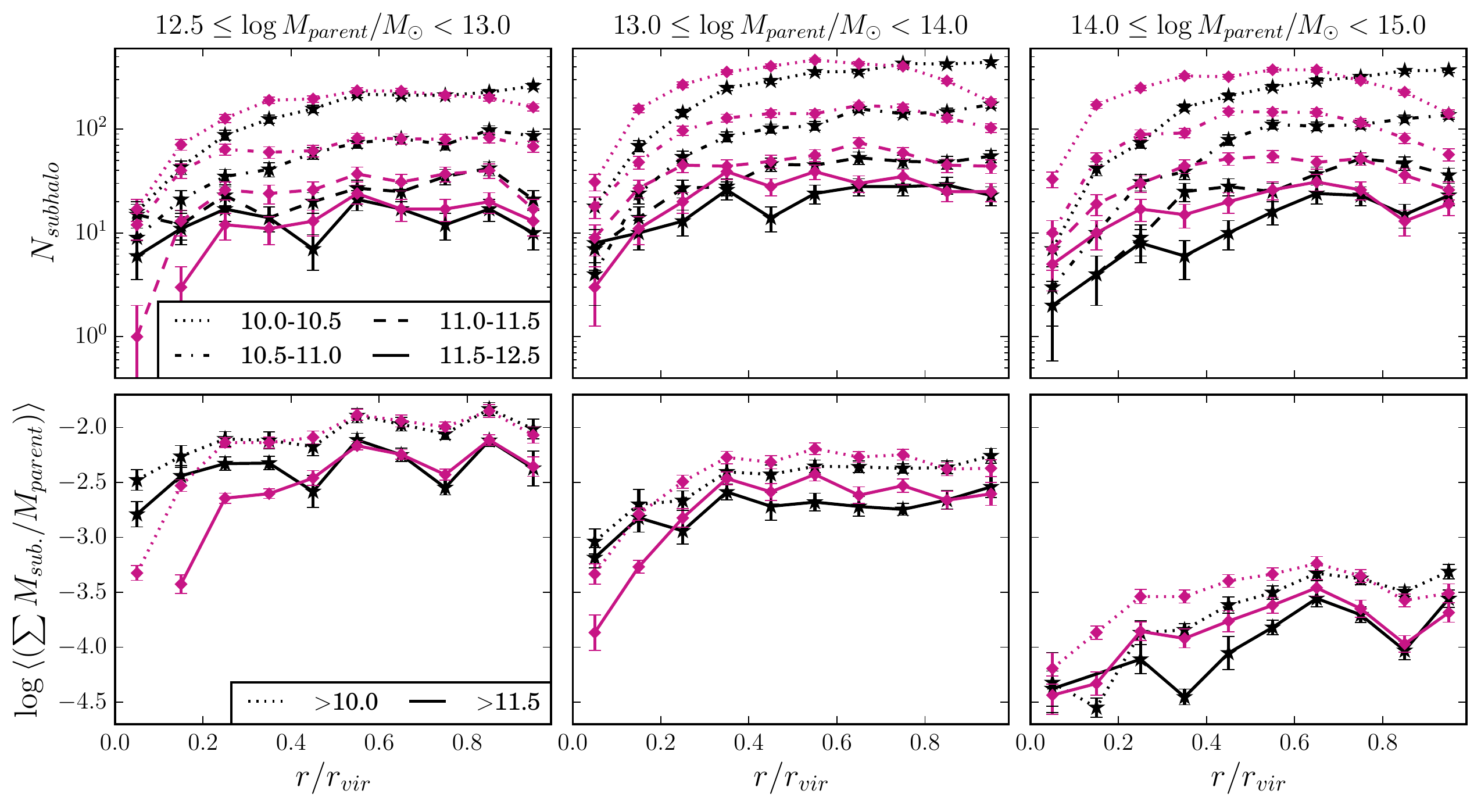}
\caption{Radial profiles of the total numbers of subhalos (\emph{Top}) and fractional subhalo mass averaged by the number of parent halos (\emph{Bottom}) as in Fig. \ref{fig:hmfRadCount}, but separated by environment. (In the bottom panel, we only show the first and last bins for clarity's sake and since the results in the intermediate bins are similar to the ones shown in Fig. \ref{fig:hmfRadCount}). The results are similar to those of the total population, with the biggest differences in total subhalos numbers seen in the most highest parent-halo-mass bin, the least in the lowest parent-halo-mass bin.} \label{fig:hmfRadCountBin}
\end{figure*}

Fig. \ref{fig:hmfRadCountBin} shows the radial profiles of the total numbers of subhalos (\emph{Top panels}) and the fractional mass in subhalos (\emph{Bottom panels}) for the highest resolution simulation as in Fig. \ref{fig:hmfRadCount} here separated into bins of parent halo mass. The different linestyles represent different bins of $\log{M_{subhalo}}$. As with the total subhalo sample, AHF finds more subhalos in the inner regions at least in the lower subhalo mass bins, although the differences are most pronounced in the highest parent mass bin. However, ROCKSTAR appears to assign more mass to subhalos in the lowest parent mass bin.

\section{Radial distributions of galaxy analogues}

\begin{figure*}
\includegraphics[width=\linewidth]{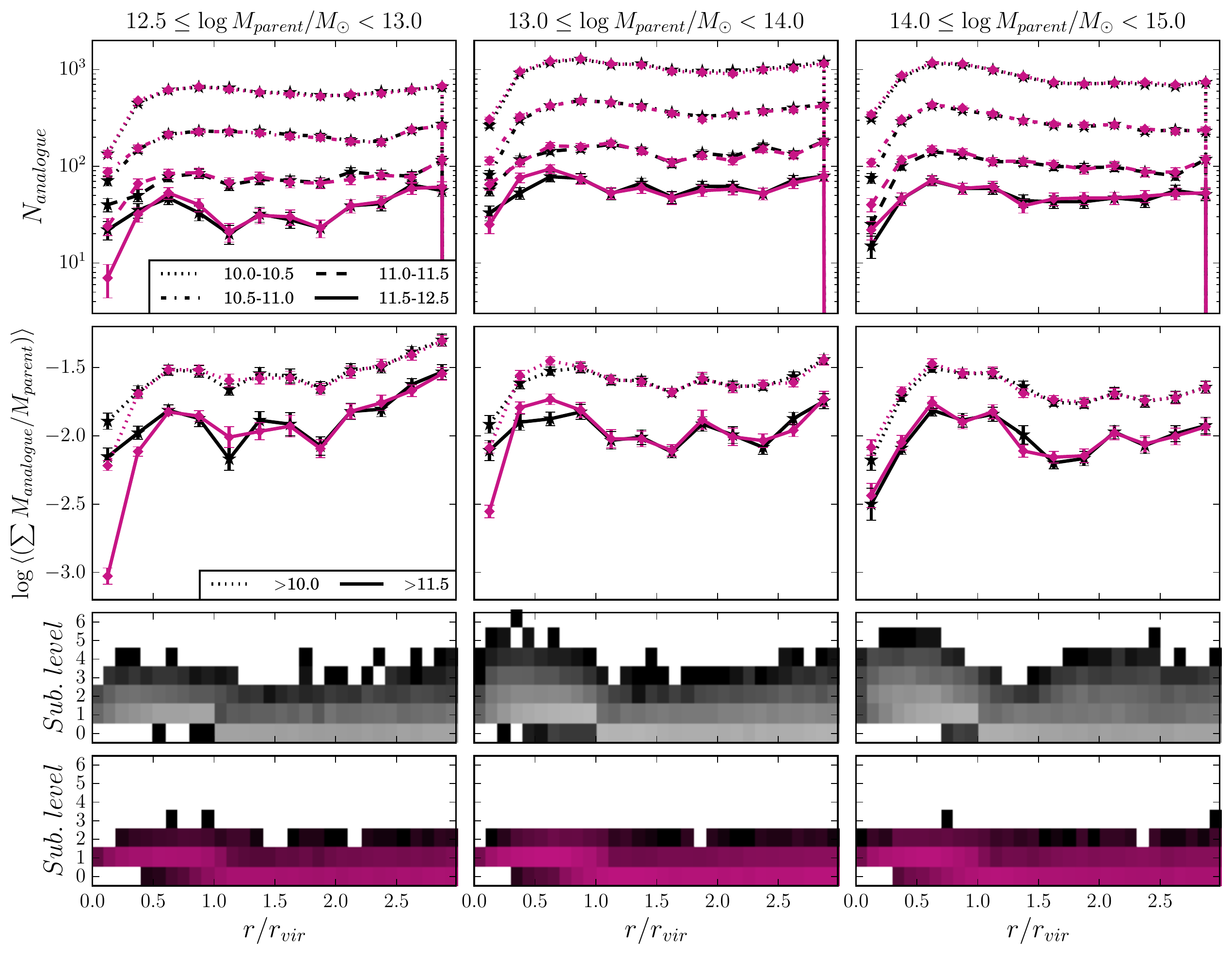}
\caption{Total numbers of analogues (\emph{Row 1}) and fractional analogue mass averaged by number of parent halos (\emph{Row 2}) vs. radial separation from the parent halo centre as in Fig. \ref{fig:hmfGACount}, but separated by environment. We only show the first and last bins of analogue mass in \emph{Row 2} for clarity. The analogue populations selected by both halo finders are consistent with each other in the lower mass bins in terms of numbers of analogues, regardless of environment. The two differ in the highest mass bin however, resulting in significant differences in terms of the total mass assigned to these analogues in the lowest parent-halo-mass bin. \emph{Rows 3 \& 4}: 2D histograms of radial separation from parent halo centre and subhalo level as in Fig. \ref{fig:hmfGACount}, but separated by environment. The analogues selected from the ROCKSTAR catalogues are at higher subhalo levels as compared to those from AHF, although expectedly, at lower levels in the lowest parent halo mass systems.} \label{fig:hmfGACountBin}
\end{figure*}

Fig. \ref{fig:hmfGACountBin} shows the radial profiles of the total numbers of subhalos (\emph{Row 1}) and the fractional mass in galaxy analogues (\emph{Row 2}) as in Fig. \ref{fig:hmfGACount}, here separated into bins of parent halo mass. Qualitatively, the results are the same as those for the total analogue sample - both halo finders find similar numbers of analogues which account for similar amounts of mass outside $\sim0.5r_{vir}$. Within $\sim0.5r_{vir}$ however, ROCKSTAR detects more analogues, especially in the lowest parent mass bin. \emph{Rows 3 \& 4} in Fig. \ref{fig:hmfGACountBin} show the subhalo levels the analogues are found at - the ROCKSTAR analogues are again found at deeper levels within the subhalo hierarchy as compared to AHF.


\bsp	
\label{lastpage}
\end{document}